\documentclass{cupbook}
\usepackage{graphicx}
\usepackage{epsfig}
\usepackage{psfig}
\usepackage{natbib}
\def\etal{et al.}

\def\teff{\ifmmode T_{\rm eff} \else $T_{\mathrm{eff}}$\fi}

\def\ltsima{$\buildrel<\over\sim$}
\def\lsim{\lower.5ex\hbox{\ltsima}}
\newcommand{\hi}{H~{\sc i}}
\newcommand{\hii}{H~{\sc ii}}
\newcommand{\ha}{\ifmmode {\rm H}\alpha \else H$\alpha$\fi}
\newcommand{\hb}{\ifmmode {\rm H}\beta \else H$\beta$\fi}
\newcommand{\lya}{\ifmmode {\rm Ly}\alpha \else Ly$\alpha$\fi}

\newcommand{\heii}{He~{\sc ii}}
\newcommand{\Heiiuv}{He~{\sc ii} $\lambda$1640}

\def\micron{$\mu$m}
\def\kms{km s$^{-1}$}

\def\msun{\ifmmode M_{\odot} \else M$_{\odot}$\fi}
\def\msunyr{\ifmmode M_{\odot} {\rm yr}^{-1} \else M$_{\odot}$ yr$^{-1}$\fi}
\def\zsun{\ifmmode Z_{\odot} \else Z$_{\odot}$\fi}
\def\zcrit{\ifmmode Z_{\rm crit} \else Z$_{\rm crit}$\fi}
\def\lsun{\ifmmode L_{\odot} \else L$_{\odot}$\fi}

\def\mup{\ifmmode M_{\rm up} \else M$_{\rm up}$\fi}
\def\mlow{\ifmmode M_{\rm low} \else M$_{\rm low}$\fi}
\def\lbol{\ifmmode L_{\rm bol} \else L$_{\rm bol}$\fi}

\def\aap{A\&A}

\def\aj{AJ}
\def\apj{ApJ}
\def\apjl{ApJL}

\def\mnras{MNRAS}
\def\pasp{PASP}
\newcommand{\oh}{\ifmmode 12 + \log({\rm O/H}) \else$12 + \log({\rm
O/H})$\fi}
\def\flyf{\ifmmode f_{\rm Lyf} \else $f_{\rm Lyf}$\fi}
\def\pz{\ifmmode P(z) \else $P(z)$\fi}
\def\ki2{\ifmmode \chi^2 \else $\chi^2$\fi}
\def\zphot{\ifmmode z_{\rm phot} \else $z_{\rm phot}$\fi}
\def\zfit{\ifmmode z_{\rm fit} \else $z_{\rm fit}$\fi}

\newcommand{\vexp}{\ifmmode v_{\rm exp} \else $v_{\rm exp}$\fi}
\newcommand{\nh}{\ifmmode N_{\rm H} \else $N_{\rm H}$\fi}
\newcommand{\la}{\raisebox{-0.5ex}{$\,\stackrel{<}{\scriptstyle\sim}\,$}}
\newcommand{\ga}{\raisebox{-0.5ex}{$\,\stackrel{>}{\scriptstyle\sim}\,$}}
\title{\it Primeval galaxies\footnote{To appear in: 
``The emission line Universe'', XVIII Canary Islands
Winter School of Astrophysics, Ed.\ J. Cepa, Cambridge Univ.\ Press}}
\author[Daniel Schaerer]{Daniel Schaerer \\
Observatoire de Gen\`eve \\ Universit\'e de Gen\`eve\\
51, chemin des Maillettes \\
CH-1290 Sauverny \\ Switzerland \\ (daniel.schaerer@obs.unige.ch) \\
{\rm and}\\
Observatoire Midi-Pyr\'{e}n\'{e}es \\
 Laboratoire
           d`Astrophysique, UMR 5572 \\ 
14 Avenue E.Belin \\
 F-31400             Toulouse, France}

\begin{document}
\pagenumbering{roman}
\maketitle
\tableofcontents
%\cleardoublepage
\pagenumbering{arabic}

\chapter[Primeval galaxies]{\it Primeval galaxies}

%%%%%%%%%%%%%%%%%%%%%%%%%%%%%%%%%%%%%%%%%%%%%%%%%%%%%%%%%%%%%%%%%%%%%%%%
\section{Introduction}
What do we mean by primordial?
According to the Webster dictionary ``Primeval: adj.\ 
 	[primaevus, from: primus first + aevum age]
of or relating to the earliest ages (as of the world or human history)''.
In these lectures we will follow this definition and mostly
discuss topics related to galaxies in the ``early'' universe, whose limit we
somewhat arbitrarily define at redshifts $z \ga $ 6, corresponding approximately to 
the first billion years (Gyr) after the Big Bang.
In contrast the frequently employed adjective ``primordial'', defined as
``Primordial: adj. 
  [primordialis, from primordium origin, from primus first + ordiri to begin]
a) first created or developed 
b) existing in or persisting from the beginning (as of a solar system or universe) 
c) earliest formed in the growth of an individual or organ'',
should not be used synonymously, for obvious reasons.
Luckily ``primeval'' encompasses more than ``primordial'', otherwise there would
not be much observational aspects to discuss (now in 2006-2007) in these
lectures!

If we follow the history of discoveries of quasars and galaxies over
the last few decades it is indeed impressive to see how progress has
been made in detecting ever more distant objects, increasing samples
at a given redshift, and in their analysis and interpretation.  During
the last decade, approximately since the pioneering observations of
the Hubble Deep Field in 1996 (Williams \etal\ 1996) and the
spectroscopic studies of a large sample of star forming galaxies at
redshift 3 by Steidel and collaborators (Steidel \etal\ 1996), the
observational limits have continuously been pushed further reaching
now record redshifts of $z \sim$ 7 (secure, Iye \etal\ 2006) but maybe
up to $\sim 10$ (cf.\ Pell\'o \etal\ 2004, Richard \etal\ 2006, Stark
\etal\ 2007).

Most of this progress has only been possible thanks to the Hubble Space
Telescope, to the availability of 10m class telescopes (Keck, VLT, SUBARU),
and to continuous improvements in detector technologies, especially in the
optical and near-IR domain. Recently the IR Spitzer Space Telescope, with
its 60cm mirror, has begun to play an important role in characterising
the properties of the highest redshift galaxies.

Not only have observations progressed tremendously. Theory and numerical
simulations now provide very powerful tools and great insight into
the physics in the early Universe, first stars and galaxies.
Within the model of hierarchical structure formation we have the following
simplified global picture of primeval galaxies, their formation and
interactions with the surrounding medium.
Schematically, following the growth of quantum fluctuations after the Big Bang,
one has in parallel:
structure formation (hierarchical),
star formation in sufficiently massive halos,
``local'' and ``global'' chemical evolution (including dust formation),
and ``local'' and ``global'' reionisation
\footnote{By {\em local} we here mean within a dark-matter (DM) halo, proto-cluster, or galaxies,
i.e.\ at scales corresponding to the interstellar medium (ISM), intra-cluster
medium (ICM), up to the ``nearby'' intergalactic medium (IGM). 
The {\em global} scale refers here to cosmic scales, i.e.\ scales
of the IGM.}.
These different processes are coupled via several feedback mechanisms 
(radiation, hydrodynamics).
In this way the universe the first stars and galaxies are thought to form,
to begin their evolution, to contribute to the chemical enrichment and 
dust production, and to gradually reionise the universe from shortly
after the Big Bang to approximately 1 Gyr after that.

This global scenario and its various physical ingredients have been
presented in depth in several excellent reviews to which the reader
is referred to (Barkana \& Loeb 2001, Bromm \& Larson 2004,
Ciardi \& Ferrara 2005, Ferrara 2006). 
In these lectures I shall only briefly outline the most important
theoretical aspects concerning the first stars and galaxies and their
expected properties (Sect.\ \ref{s_theory}).
In Sect.\ \ref{s_lya} I will introduce and discuss \lya, one of if not 
the strongest emission line in distant star forming galaxies, and review 
numerous results concerning this line and its use as a diagnostic tool.
Finally I will present an overview of our current observational 
knowledge about distant galaxies, mostly \lya\ emitters and Lyman
break galaxies (Sect.\ \ref{s_dist}).
Open questions and some perspectives for the future are discussed
in Sect.\ \ref{s_future}.
It is the hope that these lectures may be helpful for students and
other researchers in acquiring an overview of this very active and
rapidly changing field, basics for its understanding, and maybe
also provide some stimulations for persons working on related topics
to explore the rich connections between different fields intertwined
in the early universe and contributing to the richness of astrophysics.

%%%%%%%%%%%%%%%%%%%%%%%%%%%%%%%%%%%%%%%%%%%%%%%%%%%%%%%%%%%%%%%%%%%%%%%%
\section{PopIII stars and galaxies: a ``top-down'' theoretical approach}
\label{s_theory}
We shall now briefly summarise the expected theoretical properties
governing the first generations of stars and galaxies, i.e.\ objects
of primordial composition or very metal-poor.

\subsection{Primordial star formation}
In present-day gas, with a heavy element mass fraction (metallicity)
up to $\sim$ 2\%,  C$^+$, O, CO, and dust grains are excellent radiators
(coolants) and the thermal equilibrium timescale is much shorter than the 
dynamical timescale. Hence large gas reservoirs can cool and collapse rapidly, leading 
to coulds with typical temperature ≈of $\sim$ 10 K.
In contrast, in primordial gas cloud would evolve almost adiabatically, since 
heavy elements are absent and H and He are poor radiators for $T < 10^4$ K.
However, molecules such as H$_2$ or HD, can form and cool the gas in these
conditions. Approximately, it is found that at metallicities 
$Z \la \zcrit=10^{-5\pm 1}$\zsun, these molecules dominate the cooling
(e.g.\ Schneider \etal\ 2002, 2004).

Starting from the largest scale relevant for star formation (SF) in galaxies, 
i.e.\ the scale of the DM halo, one can consider the conditions necessary
for star formation (see e.g.\ Barkana \& Loeb 2001, Ferrara 2007). 
Such estimates usually rely on timescale arguments.
Most importantly, the necessary condition for fragmentation that the cooling
timescale is shorter than the free-fall timescale, 
$t_{\rm cool} \ll t_{\rm ff}$, translates to a minimum mass $M_{\rm crit}$
of the DM halo for SF to occur as a function of redshift.
A classical derivation of $M_{\rm crit}$ is found in Tegmark \etal\ (1997);
typical values of $M_{\rm crit}$\footnote{Remember: This denotes the total DM mass,
not the baryonic mass.} are $\sim 10^7$ to $10^9$ \msun\ 
from $z \sim$ 20 to 5. However, the value of $M_{\rm crit}$ is subject 
to uncertainties related to the precise cooling function
and to the inclusion of other physical mechanisms (e.g.\
ultra-high energy cosmic rays), as discussed e.g.\ in the
review of Ciardi \& Ferrara (2005).

After SF has started within a DM halo, the ``final products'' may be quite
diverse, depending in particular strongly on a variety of radiative and
mechanical feedback processes. 
Schematically, taking fragmentation and feedback into account, 
one may foresee the following classes of objects
according to Ciardi \etal\ (2000):
``normal'' gaseous galaxies, naked star clusters (i.e.\ ``proto-galaxies'' which
have blown away all their gas), and dark objects (where no stars
formed, or where SF was rapidly turned off due negative radiative feedback).
At very high redshift ($z > 10$) naked star clusters may be 
more numerous than gaseous galaxies.

How does SF proceed within such a small ``proto-galaxy'' and what stars
will be formed? 
Fragmentation may continue down to smaller scales. In general 
the mass of the resulting stars will depend on the fragment mass,
the accretion rate, radiation pressure, and other effects such
as rotation, outflows, competitive accretion etc., forming a rich
physics which cannot be described here (see e.g.\ reviews by
Bromm \& Larson 2004, Ciardi \& Ferrara 2005 and references therein)
Most recent numerical simulations following early star formation
at very low metallicities agree that at $Z \la \zcrit$ the smallest
fragment are quite massive, and that they undergo a runaway collapse
accompanied with a high accretion rate resulting in (very) massive stars
(10--100 \msun\ or larger), compared to a typical mass scale 
of $\sim 1 \msun$ at ``normal'' (higher) metallicities
(cf.\ Bromm \& Larson 2004). 
This suggests that the stellar initial mass function (IMF) may differ
significantly from the present-day distribution at 
$Z \la \zcrit=10^{-5\pm 1} \zsun$.
The value of the critical metallicity is found to be determined
mostly by fragmentation physics; in the transition regime
around \zcrit\ the latter may in particular also depend on dust properties
(cf.\ Schneider \etal\ 2002, 2004).

Determining the IMF at $Z < \zcrit$ observationally is difficult and
relies mostly on indirect constraints (see e.g.\ Schneider \etal\
2006). The most direct approaches use the most metal poor Galactic
halo stars found.  From counts (metallicity distributions) of these
stars Hernandez \& Ferrara (2001) find indications for a increase of
the characteristic stellar mass at very low $Z$.  Similar results have
been obtained by Tumlinson (2006), using also stellar abundance
pattern. However, no signs of very massive ($> 130$ \msun) stars
giving rise to pair instability supernovae (see Sect.\ \ref{s_pisn})
have been found yet (cf.\ Tumlinson 2006).  In Sect.\ \ref{s_dist} we
will discuss attempts to detect PopIII stars and to constrain their
IMF {\em in situ} in high redshift galaxies.

\subsection{Primordial stars: properties}
Now that we have formed individual (massive) stars at low metallicity,
what are their internal and evolutionary properties?
Basically these stars differ on two main points from their
normal metallicity equivalents: the initial source of
nuclear burning and the opacity in their outer parts.
%lack of opacity due to metals
Indeed, since PopIII stars (or more precisely stars with metallicities $Z\la 10^{-9}
= 10^{-7.3} \zsun$) cannot burn on the CNO cycle like normal massive stars, their
energy production has to rely initially on the less efficient
p-p chain. Therefore these stars have higher central temperatures.
Under these conditions ($T \ga 10^{8.1}$ K) and after the build-up
of some amount of He, the 3-$\alpha$ reaction becomes possible,
leading to the production of some amounts of C.
In this way the star can then ``switch'' to the more efficient
CNO cycle for the rest of H-burning, and its structure
(convective interior, radiative envelope) is then similar
to ``normal'' massive stars.
Given the high central temperature and the low opacity (dominated by
electron scattering throughout the entire star due the lack of metals),
these stars are more compact than their PopII and I correspondents.
Their effective temperatures are therefore considerably higher,
reaching up to $\sim 10^5$ K for $M \ga 100$ \msun\ (cf.\
Schaerer 2002).
The lifetimes of PopIII stars are ``normal'' (i.e.\ $\sim$ 3 Myr at minimum),
since $L \sim M$, i.e.\ since the luminosity increases approximately linearly with
the increase of the fuel reservoir.
Other properties of ``canonical'' PopIII stellar evolution models are 
discussed in detail in Marigo \etal\ (2001), Schaerer (2002), and 
references therein.

\begin{figure}[tb]
%\centerline{\psfig{file=plot_nlyc_uv_z.eps,width=6cm}
%  \psfig{file=MS2980f5.eps,width=6cm}}
\centerline{\psfig{file=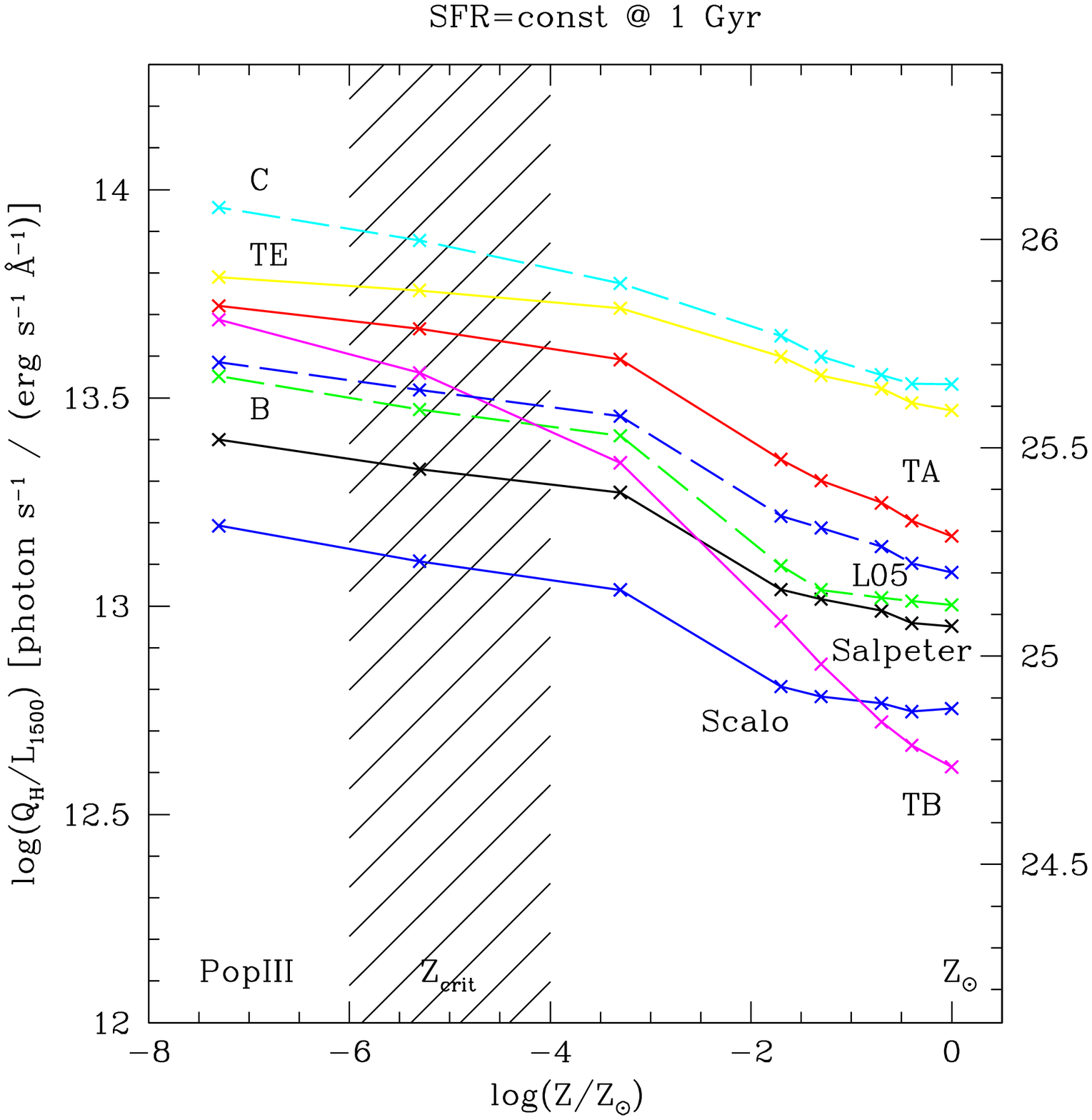,width=6cm}
  \psfig{file=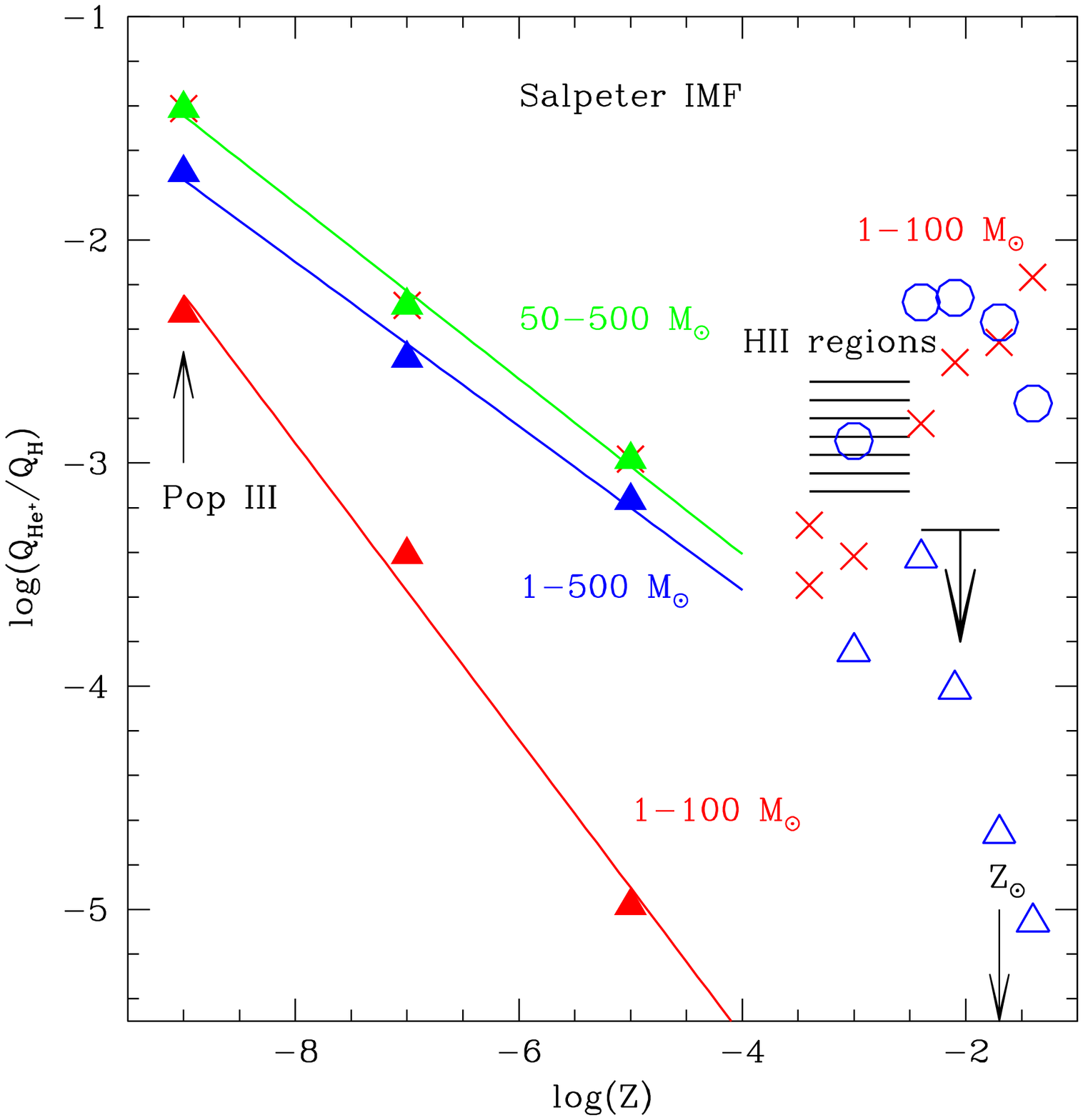,width=6cm}}
\caption{{\bf Left:} Relative output of hydrogen ionising photons to UV
    light, measured at 1500 \AA\ restframe, $Q_H/L_{1500}$, as a
    function of metallicity for constant star formation over 1 Gyr.
%$Q_H/L_{1500}$
%    is given in $L_\lambda$ units on the left side of each panel, and
%    in $L_\nu$ units on the right.  
    Results for different IMFs, including a Salpeter, Scalo and more
    top-heavy cases, are shown using different color codes.  The
    shaded area indicates the critical metallicity range where the IMF
    is expected to change from a ``normal'' Salpeter-like regime to a
    more massive IMF.  {\bf Right:} Hardness $Q({\rm He^+})/Q({\rm
      H})$ of the He$^+$ ionising flux for constant star formation as
    a function of metallicity (in mass fraction) and for different
    IMFs. At metallicities above $Z \ge 4. \, 10^{-4}$ the
    predictions from our models (crosses), as well as those of
    Leitherer \etal\ (1999, open circles), and Smith \etal\ (2002,
    open triangles) are plotted.  The shaded area and the upper limit
    (at higher $Z$) indicates the range of the empirical hardness
    estimated from \hii\ region observations.  From Schaerer (2003)}
\label{fig_1}
\end{figure}	

More sophisticated stellar evolution models including many physical
processes related to stellar rotation are now being constructed
(cf.\ Meynet \etal\ 2006, Ekstr\"om \etal\ 2006).
Whereas before it was thought that mass loss would be 
negligible for PopIII and very metal-poor stars
(since radiation pressure is very low and pulsational instabilities
may only occur during a very short phase; cf.\ Kudritzki 2002,
Baraffe \etal\ 2001), fast rotation -- due to fast initial
rotation and inefficient transport of angular momentum --
may lead to mechanical mass loss, when these stars
reach critical (break-up) velocity.
Rotation also alters the detailed chemical yields,
may lead to an evolution at hotter \teff, even to WR stars,
and it may alter final fate of PopIII/very metal-poor stars,
which may in this way even avoid the ``classical'' 
pair instability supernova (PISN, cf.\ below).
Many details and the implications of these models on 
observable properties of metal-free/very metal-poor
populations still remain to be worked out.

\subsection{Primordial stars \& galaxies: observable properties}
\label{s_obsprop}
The observable properties of individual PopIII and metal-poor
stars and of an integrated population of such stars can be predicted 
using stellar evolution models, appropriate non-LTE
stellar atmospheres, and using evolutionary synthesis
techniques (see e.g.\ Tumlinson \etal\ 2001, Bromm \etal\ 2001,
and detailed discussions in Schaerer 2002, 2003).

\begin{figure}[tb]
%\centerline{\psfig{file=plot_sed_zams.eps,width=6cm}
%  \psfig{file=plot_beta_slope.eps,width=6cm}}
\centerline{\psfig{file=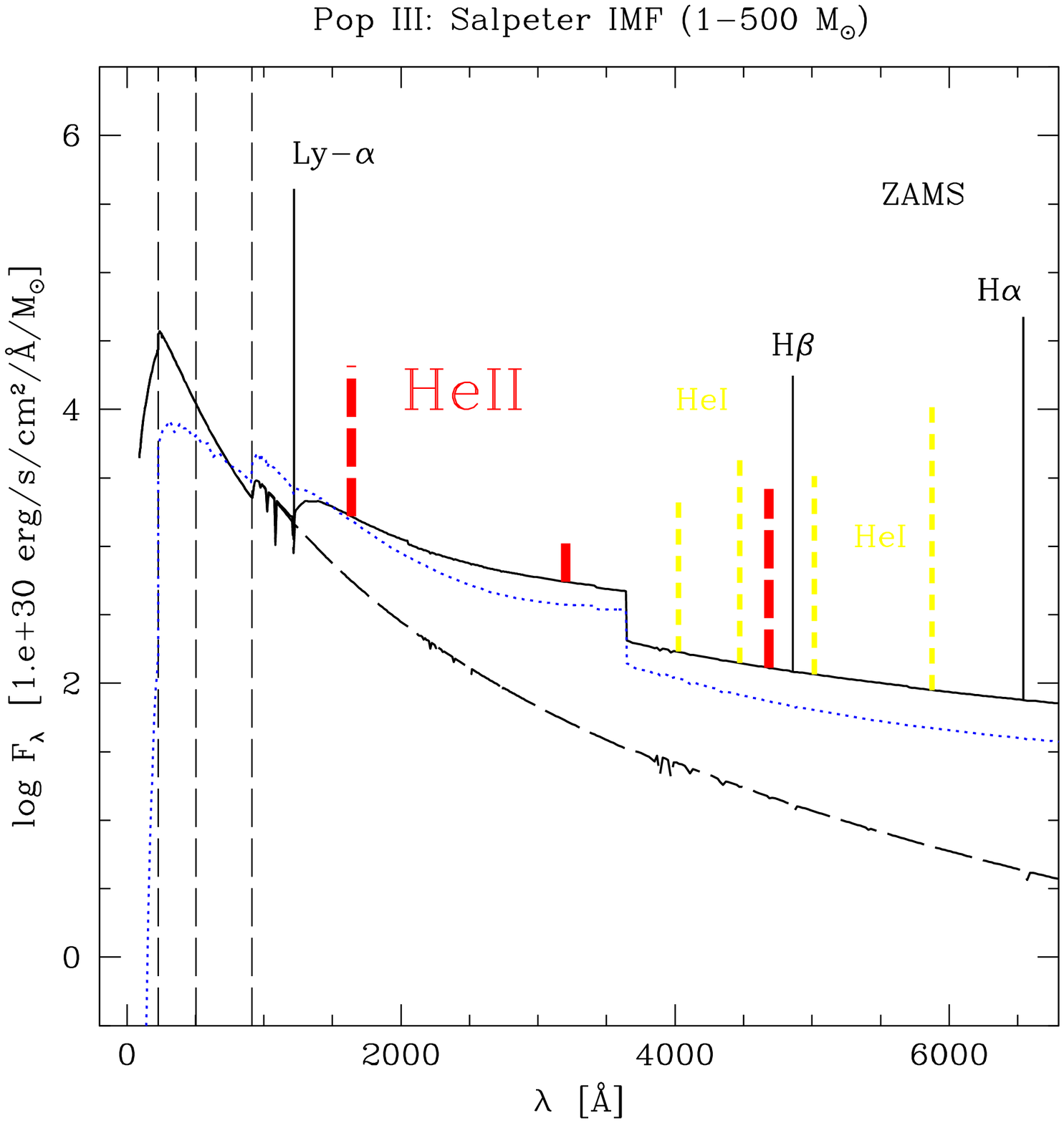,width=6cm}
  \psfig{file=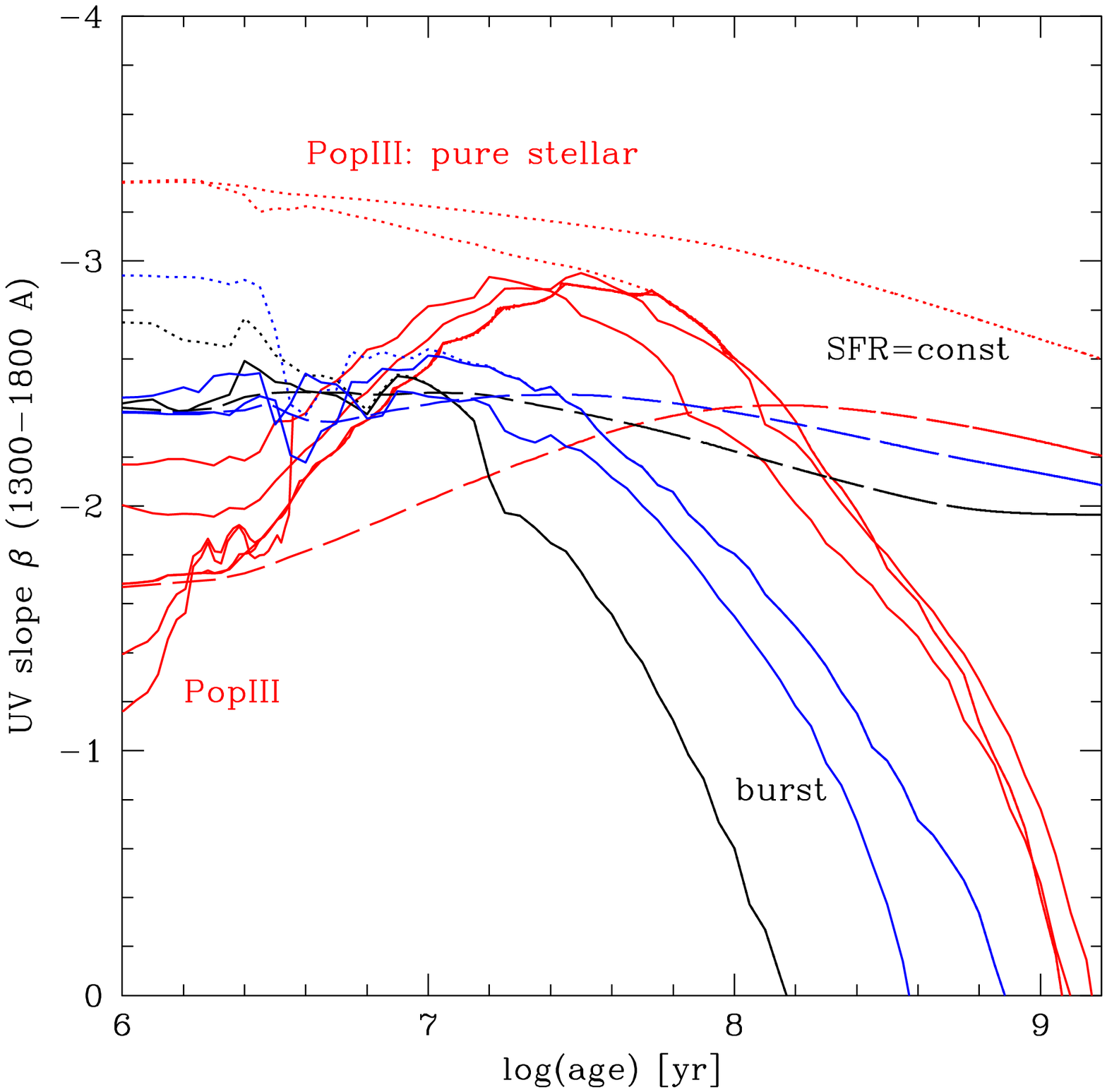,width=6cm}}
\caption{{\bf Left:} Spectral energy distribution of a very young
  PopIII galaxy including H and He recombination lines The pure
  stellar continuum (neglecting nebular emission) is shown by the
  dashed line. For comparison the SED of the $Z=1/50 \zsun$ population
  (model ZL: Salpeter IMF from 1 -- 150 \msun) is shown by the dotted
  line.  The vertical dashed lines indicate the ionisation potentials
  of H, He$^0$, and He$^+$.  Note the presence of the unique \heii\
  features (shown as thick dashed lines) and the importance of nebular
  continuous emission.  From Schaerer (2002).  {\bf Right:} Temporal
  evolution of the UV slope $\beta$ measured between 1300 and 1800
  \AA\ from synthesis models of different metallicities and for
  instantaneous bursts (solid lines) and constant SF (long dashed
  lines).  Black lines show solar metallicity models, red lines
  metallicities between $Z = 10^{-5}$ and zero (PopIII), blue lines
  intermediate cases of $Z=0.004$ and 0.0004.  The dotted lines show
  $\beta$ if nebular continuous emission is neglected, i.e.\ assuming
  pure stellar emission.  Note especially the strong degeneracies of
  $\beta$ in age and metallicity for bursts, the insensitivity of
  $\beta$ on $Z$ for constant SF, and the rather red slope for young
  very metal-poor bursts.  From Schaerer \& Pell\'o (2005).}
\label{fig_2}
\end{figure}	

Given the exceptionally high effective temperatures of PopIII stars
on the zero age main sequence, such objects emit a larger fraction
of the luminosity in the Lyman continuum and have a much harder
ionising spectrum than higher metallicity stars. E.g.\ a PopIII
star of  5 \msun\ is still an ionising source!
In other words, stellar populations at low metallicity are characterised
by a high ionisation efficiency (per unit stellar mass formed)
and by a hard spectrum, as illustrated in Figure \ref{fig_1}.
For a unchanged IMF, e.g.\ Salpeter, the ionising output normalised 
to the UV flux density increases by a factor $\sim$ 2 or more from
solar metallicity to PopIII. However, this increase may be
much more substantial if the IMF favours massive stars at
low $Z$, as argued before.

The predicted integrated spectrum of a very young (ZAMS) ensemble of 
PopIII stars is shown in Fig.\ \ref{fig_2}. Its main characteristics
are the presence of strong H emission lines (in particular
strong \lya, cf.\ below) due to the strong ionising flux, 
He$^+$ recombination lines (especially \Heiiuv) due to spectral 
hardness, and strong/dominating nebular continuum emission
(cf.\ Schaerer 2002).
The strength of \lya\ can be used to identify interesting
PopIII or very-metal poor galaxy candidates (cf.\ Sect.\ \ref{s_lya}).
The detection of nebular \Heiiuv\ (if shown to be due to stellar
photoionisation, i.e.\ non-AGN origin) would be a very interesting
signature of primordial (or very close to) stars. Indeed,
as shown on the right of Fig.\ \ref{fig_1} very hard spectra
are only predicted at $Z \la 10^{-5 \ldots -6}$ \zsun.

It is often heard that PopIII, primeval or similar galaxies should
be distinguished by bluer colors, e.g.\ measured in the rest-frame UV,
as one would naively expect.
Although indeed the colors of stars get bluer on average with
decreasing metallicity, this is not the case anymore for the
integrated spectrum of such a population, since nebular 
continuum emission (originating from the \hii\ regions surrounding
the massive stars) may dominate the spectrum, even in the UV.
This leads to a much redder spectrum, as shown in Fig.\ \ref{fig_2} (left).
Taking this effect into account leads in fact to a non-monotonous
behaviour of the slope (color) of the UV spectrum with 
metallicity, illustrated in Fig.\ \ref{fig_2} (right). 
This fact, and the dependence of the UV slope on the star formation
history on timescales shorter than $10^8$ to $10^9$ yr, corresponding 
to 10-100 \% of the Hubble time at $z\ga 6$, show that 
the interpretation of the UV slope (or color) of primeval galaxies
must be taken with great caution.

\subsection{Final fate}
\label{s_pisn}
The end stages of very metal-poor and PopIII stars may also differ
from those at higher metallicity, with several interesting
consequences also for the observational properties of primeval galaxies.
In particular such massive stars may at the end of their evolution 
show conditions in central temperature and density, such
that the creation of electron-positron pairs occurs, leading to
an instability which will completely disrupt the star.
This phenomenon is known as pair instability supernova (PISN)
\footnote{Sometimes the term pair creation SN or pair production SN
is also used.}, and a rich literature exists about
the phenomenon and many implications. Here we shall only summarise
the main salient points and recent findings.

A recent overview of the different ``final events'' and remnants 
is found in Heger \etal\ (2003).  
PISN are thought to occur for stars with initial masses
of $M \sim$ 140--260 \msun\ a very low $Z$.
Due to their high energy and to non-negligible time dilation
which increases the duration of their ``visibility'', PISN
are potentially detectable our to very high redshift 
(see e.g.\ Weinmann \& Lilly 2005, Scannapieco \etal\ 2005, Wise \& Abel
2005).
Large amounts of gas are ejected, as the event disrupts
the star completely. Furthermore, the processed matter 
contains peculiar nucleosynthesic signatures which may
in principle be distinguished from normal SN (cf.\ below).
Finally PISN are also thought to be the first dust production
factories in the universe (cf.\ Schneider \etal\ 2004).
Thus PISN may be observable directly and indirectly, which would
be very important to confirm or infirm the existence
of such massive stars, i.e.\ to constrain the IMF of the
first stellar generations.
Currently, however, there is no such confirmation, as we will
just discuss.

\subsection{Nucleosynthesis \& abundance pattern}
Among the particularities of PISN are the production of large quantities
of O and Si, which translate e.g.\ in large O/C and Si/C abundance ratios
potentially measurable in the IGM. More generally one expects:
roughly solar abundance of even nuclear charge nuclei (Si, S, Ar $\ldots$) and
deficiencies in odd nuclei (Na, Al, P, V $\ldots$) i.e.\ a strong 
so-called odd/even effect, and no elements heavier than Zn, due to 
the lack of s-and r-processes (see Heger \& Woosley 2002 for recent predictions).

Abundance studies of the most metal-poor halo stars in the Galaxy
do not show the odd/even effect predicted for PISN. In face of our
current knowledge, in particular on nucleosynthesis, quantitative
analysis of the observed abundance pattern thus disfavour IMFs with
a large fraction of stars with masses $M \sim$ 140--260 \msun\
(Tumlinson 2006). However, 
the abundance pattern and other constraints are compatible with
a qualitative change of the IMF at $Z \la 10^{-4} \zsun$ as suggested
by simulations (cf.\ above).

\subsection{Dust at high-z}
\label{s_dust}
Dust is known to be present out to the highest redshifts from 
damped \lya\ absorbers (DLA), from sub-mm emission in $z \sim 6$ Quasars
(e.g.\ Walter \etal\ 2003), from a GRB host galaxy at $z=6.3$ 
(Stratta \etal\ 2007), and possibly also from the spectral energy 
distribution (SED) of some normal galaxies at $z \sim 6$ 
(Schaerer \& Pell\'o 2005).
%Bouwens \etal\ 2006). 
We also know that dust exists in the most metal-poor galaxies, as testified
e.g.\ by the nearby galaxy SBS 0335-052 with a metallicity of $\sim 1/50$ \zsun.

Since the age of the universe at $z>6$ is $\sim$ 1 Gyr at most,
longer-lived stars cannot be invoked to explain the dust
production in primeval galaxies. Among the possible 
``short-lived'' dust producers are SNII, PISN, maybe also
Wolf-Rayet stars or massive AGB stars.
SNII are known dust producers (e.g.\ SN1987A), although maybe
not producing enough dust. Efficient dust production is found 
in explosions of SNII and PISN (e.g.\ Todini \& Ferrara 2001, 
Schneider \etal\ 2004). At zero metallicity PISN may provide 
a very very efficient mechanism, converting up to 7-20\% of PISN 
mass into dust. 

Evidence for dust produced by SN has been found from 
the peculiar extinction curve in the BAL QSO SDSS1048+46
at $z=6.2$, which shows good agreement with SN dust models
(Maiolino \etal\ 2004). 
Similar indications have been obtained recently from a
GRB host galaxy at $z6.3$ (Stratta \etal\ 2007). 
If this is a general feature remains, however, to be established.
Furthermore the most important questions, including 
how common is dust in high-z galaxies, in what quantities,
up to which redshift, etc.\ have not yet been touched.
Forthcoming IR to sub-mm facilities such as Herschel and especially
ALMA will allow to address these important issues.

%%%%%%%%%%%%%%%%%%%%%%%%%%%%%%%%%%%%%%%%%%%%%%%%%%%%%%%%%%%%%%%%%%%%%%%%
\section{\lya\ physics and astrophysics}
\label{s_lya}
As \lya, one of the strongest emission lines in the UV,
plays an important role in searches for and studies of 
distant and primeval galaxies, we wish to devote one lecture to
this line, the basic principles governing it, its 
diagnostics and possible difficulties, empirical findings etc.
To the best of my knowledge few if any reviews or lectures
summarising these topics in a single text exist. 

\subsection{ISM emission and ``escape''}
All ionised regions, i.e.\ \hii\ regions, the diffuse ISM and alike regions
in galaxies, emit numerous emission lines including recombination lines 
from H, He, and other atoms, and forbidden, semi-forbidden, and fine structure
metal lines resulting from deexcitations of these atoms
(see the textbooks of Osterbrock \& Ferland 2006 or Dopita \& Sutherland
2003, and Stasi\'nska in these lecture notes).
All galaxies with ongoing massive star formation
(somewhat loosely called ``starbursts'' hereafter)
emitting intense UV radiation and an ionising flux
(i.e.\ energy at $>$ 13.6 eV) will thus ``intrinsically'',
viz.\ at least in their \hii\ regions, show \lya\ emission.

From quite simple considerations one can find that
the luminosity in a given H recombination line is proportional
to the number of ionising photons (i.e.\ Lyman-continuum photons),
$L($\lya,\ha,$\ldots)=c_l Q_H$, where $Q_H$ is the Lyman-continuum flux
in photons $s^{-1}$ and $c_l$ a ``constant'' depending somewhat on the
nebular temperature $T_e$ and the electron density $n_e$. 
For hydrogen, $\sim 2/3$ of the recombinations lead to the emission
of a \lya\ photon, corresponding to the transition from level 2 to the 
ground state (cf.\ Spitzer 1978, Osterbrock \& Ferland 2006).
Furthermore the relative intensities of two different H recombination
lines are known and relatively slowly varying functions of temperature
and density, e.g.\ $I(\lya)/I(Hn)=c(T,n_e)$. 

Already in the sixties it was recognised that \lya\ could be
import for searches of primeval galaxies (e.g.\ 
Partidge \& Peebles 1967). Indeed, at (very) low
metallicities the \lya\ line is expected to be strong if not
dominant for several reasons:
an increasing ionising flux from stellar populations,
\lya\ can become the dominant cooling line when few metals
are present,
an increased emissivity due to collisional excitation
in a nebula with higher temperature.
As a result up to $\sim 10$\% of the bolometric luminosity
may be emitted in \lya, rendering the line potentially detectable 
out to the highest redshifts!
This prospect triggered various searches for distant 
\lya\ emitters, which remained, however, basically unsuccessful
until the 1990ies (see Sect.\ \ref{s_dist}), for the
reasons discussed below.
In any case, it is interesting to note that most of the observational
features predicted nowadays for PopIII galaxies (cf.\ Sect.\ \ref{s_obsprop})
were anticipated by early calculations, such as Partridge \& Peebles' (1967),
including of course the now famous Lyman-break (Sect.\ \ref{s_dist}).

To anticipate it is useful to mention already here the basics of the
\lya\ escape problem.
In short, 
%galaxies are in general optically thin in the observable UV ($>912$ \AA).
%However, very rapidly the \lya\ line becomes optically thick (for column 
%densities $N_{HI} \ga 10^{13}$ cm$^{-2}$). 
for very low column densities of $N_{HI} \ga 10^{13}$ cm$^{-2}$
the \lya\ line becomes already optically thick.
Therefore radiation transfer within the galaxy determines the emergent line
profile and the \lya\ ``transmission''!
Furthermore, dust may destroy \lya\ photons. 
Overall, the fate of \lya\ photons emitted in a galaxy can be one of the following:
{\em 1)} scattering until escape forming thus an extended \lya\ ``halo'';
{\em 2)} destruction by dust; or
{\em 3)} destruction through 2 photon emission. However, this process is only 
possible in the ionised region.

\subsection{\lya: the observational « problem »}
As already mentioned, there were several unsuccessful searches for \lya\ emission
from $z \sim$2--3  ``primordial'' galaxies in the 1980-1990ies
(cf.\ Pritchet 1994).
Why these difficulties occurred could be understood by observations of nearby 
starbursts, which found one or two puzzles, namely 
a small number of \lya\ emitting galaxies and/or lower than expected \lya\ emission.
The second puzzle could of course in principle explain the first one.
In particular UV spectra of nearby starbursts (\lya) taken with the IUE satellite
and optical spectra (\ha, \hb) showed that:
{\em i)} after extinction correction,
the relative line intensity of e.g.\ $I(\lya)/I(\hb)$ was much smaller than
the expected case B value and the \lya\ equivalent width $W(\lya)$ smaller
than expected from evolutionary synthesis models, and
{\em ii)} these findings do not depend on metallicity 
(e.g.\ Meier \& Terlevich 1981, %
Hartmann \etal\ 1984, 
Deharveng \etal\ 1986,
and other later papers).
%$\ldots$, until Giavalisco \etal\ 1996).

Among the possible explanations put forward were:
{\em a)} Dust which would destroy the \lya\ photons (cf.\ Charlot \& Fall 1993).
{\em b)} An inhomogeneous ISM geometry, not dust, as a primarily 
determining factor (Giavalisco \etal\ 1996).
{\em c)} A short ``duty cycle'' of SF to explain the small number of \lya\ emitters.
{\em d)} Valls-Gabaud (1993) argued that with an appropriate, i.e.\ metallicity-dependent,
extinction law (i) was no problem. Also, he stressed the importance of
underlying stellar \lya\ absorption.

Rapidly dust as a sole explanation was ruled out by the observations
of I Zw 18 and SBS 0335-052, the most metal-poor stabursts known,
which show no \lya\ emission, actually even a damped \lya\ absorption
profile (Kunth \etal\ 1994, Thuan \& Izotov 1997).  However, we now
know (from ISO and Spitzer) that these objects contain also
non-negligible amounts of dust (Thuan \etal\ 1999, Wu \etal\ 2007),
although it is not clear if and how it is related to the line emitting
regions, in particular spatially.  From the absence of correlations
between different measurements of extinction, Giavalisco \etal\ (1996)
suggest that an inhomogeneous ISM geometry must be the primarily
determining factor, not dust. However, no quantification of this
effect was presented or proposed.  More detailed observations of local
starbursts have since provided new important pieces of information we
will now briefly summarise.

\begin{figure}[tb]
%\centerline{\psfig{file=hayes_etal05_fig7.eps,width=12cm}}
%\centerline{\psfig{file=figure5schaerer_compress.eps,angle=270,width=12cm}}
%\centerline{\psfig{file=figure5schaerer.eps,width=12cm}}
\caption{{\bf See Fig. 7 of Hayes \etal\ (2005) left out due to space 
limitations}. 
Observations of the nearby Blue Compact Galaxy ESO 338-IG04 from 
Hayes \etal\ (2005). {\bf Left:} \lya\ equivalent width map.
 Regions of high equivalent width show
 up in dark colours. Particularly visible are the diffuse emission
 regions outside the starburst region. Much local structure can be
 seen, particularly around knot A (the main UV know) and the other bright continuum
 sources.  {\bf Right:} false colour image showing [OIII] in red, the 
 UV continuum in green and the continuum
 subtracted \lya\ image in blue.}
\label{fig_hayes}
\end{figure}	

\subsection{Lessons from local starbursts}
Indeed high-dispersion spectroscopy with HST has shown the presence
of neutral gas outflows in 4 starbursts with \lya\ in emission
(P-Cygni profiles), whereas other starbursts with broad damped
\lya\ absorption do not show velocity shifts between the ionised
emitting gas and the neutral ISM traced by O~{\sc i} or
Si~{\sc ii} (Kunth \etal\ 1998).
The metallicities of these objects range from $12+\log($O/H$) \sim$
8.0 to solar, their extinction is $E_{B-V} \sim$ 0.1--0.55.
From these observations Kunth \etal\ (1998) suggest that outflows
and superwinds are the main determining factor for \lya\ escape.

2-3 D studies of \lya\ and related properties in nearby starbursts
have been carried out with HST (UV) and integral field spectroscopy
(optical) to analyse at {\em high spatial resolution} the distribution
and properties of the relevant components determining \lya, i.e.\ the
young stellar populations, their UV slope (a measurement of the
extinction), the ionised gas, and the resulting \lya\ emission,
absorption and the local line profile (e.g.\ Mas-Hesse \etal\ 2003,
Kunth \etal\ 2003, Hayes \etal\ 2005).
In ESO 338-IG04 (Tol 1914-416), for example, diffuse \lya\ is observed
corresponding to $\sim 2/3$ of the total flux observed in large apertures
(e.g.\ IUE), confirming thus the existence of a \lya\ resonant scattering halo
(Hayes \etal\ 2005).
No clear spatial correlation between stellar ages and \lya\ is found.
However, correlations between the \lya\ line kinematics and other kinematic
tracers (NaID or \ha) are found.
%* different regions: different H kinematics

Another interesting case is ESO 350-IG038, where Kunth \etal\ (2003)
find two young star forming knots (B and C) with similar, high
extinction, one showing \lya\ emission the other not. Hence dust
absorption cannot be the dominant mechanism here. Based on the
observed \ha\ velocity field, Kunth \etal\ suggests that kinematics is
primarily responsible for the observed differences between the two
regions.

A ``unifying'' scenario to explain the observed diversity of \lya\
profiles in terms of an evolutionary sequence of starburst driven 
supershells/superwind has been presented by Tenorio-Tagle \etal\ 
(1999) and confronted with local observations in the same
paper and more in depth by Mas-Hesse \etal\ (2003).

In short we retain the following empirical results from nearby
starbursts on \lya: $W(\lya)$ and \lya/\hb\ are often smaller than the
case B prediction.  No clear correlation of \lya\ with metallicity,
dust, and other parameters is found.  Strong variations of \lya\ are
observed within a galaxy.  A \lya\ scattering ``halo'' is observed.
Starbursts show complex structure (super star clusters plus diffuse
ISM), and outflows are ubiquitous.
From the various observations it is clear that the formation of
\lya\ is affected by:
{\em 1)} ISM kinematics, {\em 2)} ISM (HI) geometry, and {\em 3}
dust. However, the precise order of importance remains unclear
and may well vary between different objects.

New, more complete high spatial resolution observations are needed.
In parallel quantitative modeling including known constraints (stars,
emitting gas, HI, dust plus kinematics) with 3D radiation transfer
model remains to be done.

%but no constraint on HI kinematics at this spatial scale (requires SKA)!

\subsection{\lya\ radiation transfer}
%\subsection{\lya\ radiation transfer w/o dust}
\subsubsection{Basic line formation processes and examples}
To gain insight on the physical processes affecting \lya, to
understand the variety of observed line profiles and their nature, and
hence to develop quantitative diagnostics using \lya, it is important
to understand the basics of \lya\ radiation radiation transfer. To do
so we rely on the recent paper by Verhamme \etal\ (2006), where more
details and numerous references to earlier papers can be found.
Among recent papers shedding new light on \lya\ radiation transfer
we mention here the work of Hansen \& Oh (2006) and Dijkstra \etal\
(2006ab).

\begin{figure}[tb]
%\centerline{\psfig{file=5554fg01.eps,width=6cm}
%  \psfig{file=dip_t4.eps,width=6cm}}
\centerline{\psfig{file=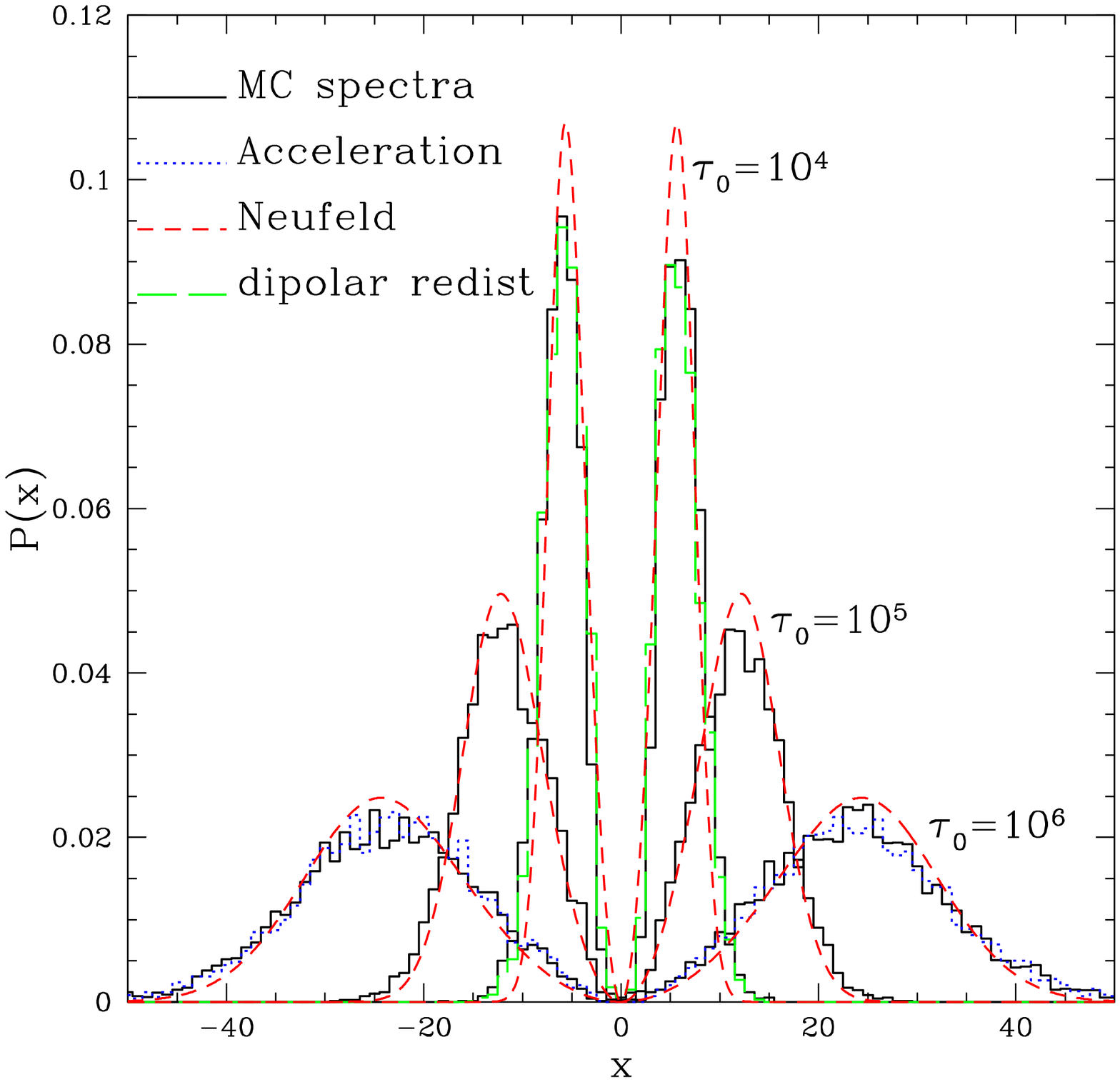,width=6cm}
  \psfig{file=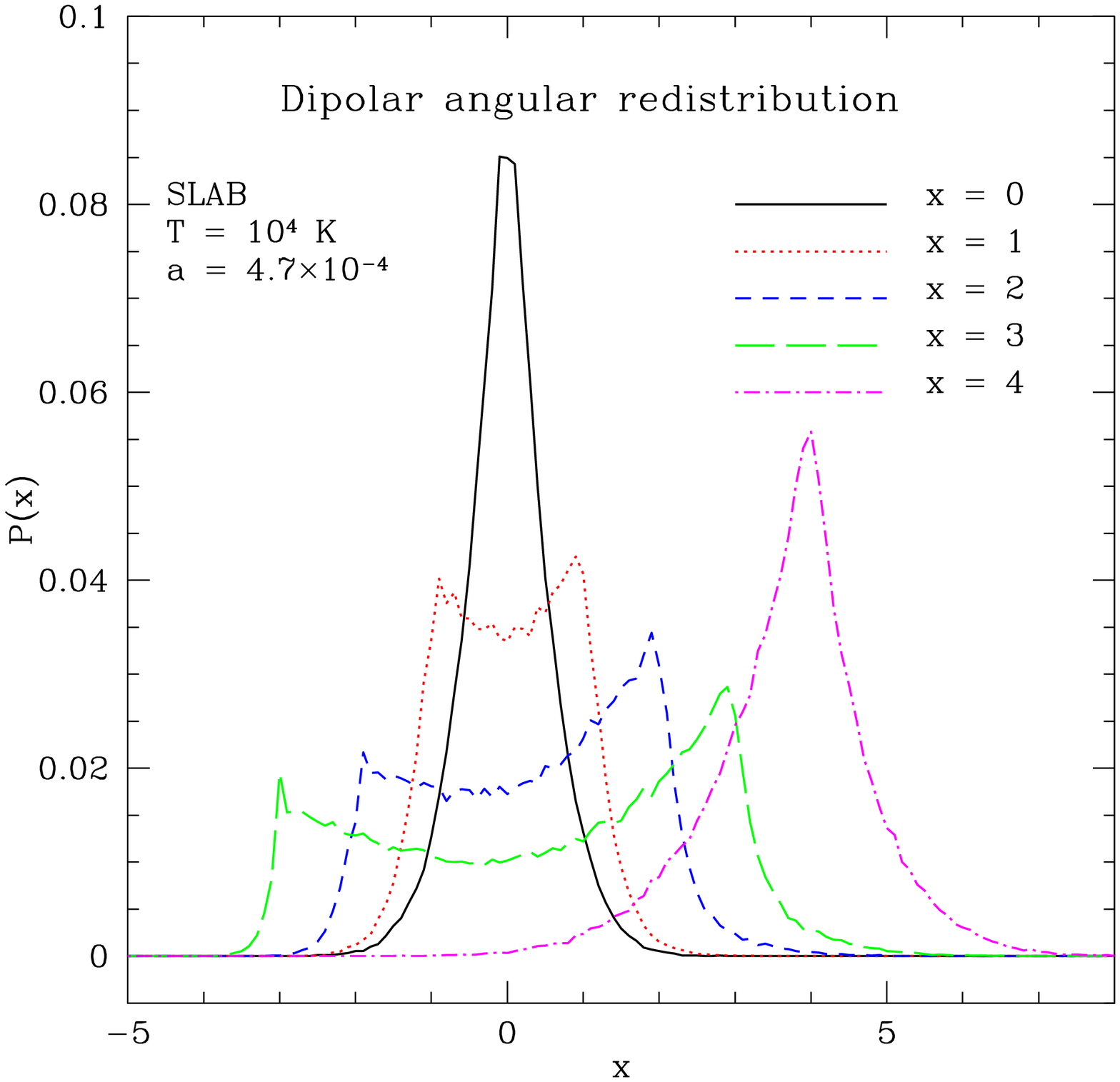,width=6cm}}
\caption{{\bf Left:} Predicted \lya\ line profile for a monochromatic
source embedded in a static medium with different \nh\ column densities.
Note the characteristic symmetric double peak profile. The separation
between the two peaks depends in particular on the total
optical depth, i.e.\ on \nh.
{\bf Right:}
Angle averaged frequency redistribution function
for specific conditions ($T$, and Voigt-parameter $a$). 
Shown is the probability distribution function for different input
frequencies $x=0$ (line center) to 4 (``wing'').
Figures from Verhamme \etal\ (2006).}
\label{fig_freq}
\end{figure}	

The \lya\ line optical depth can be written as 
\begin{equation}
\tau_x(s) = 
1.041\times 10^{-13}\,T_4^{\phantom{4}-1/2}\, N_H\, \frac{H(x,a)}{\sqrt\pi}
\label{eq_tau}
\end{equation}
where $T_4$ is the temperature in units of $10^4$ K, $N_H$ the 
neutral hydrogen column density, and $H(x,a)$ the Hjerting function 
describing the Voigt absorption profile.
Here $x$ describes the frequency shift in Doppler units, 
%\begin{equation}
$x = \frac{\nu - \nu_0}{\Delta \nu_D} = -\frac{V}{b},$
%\label{eq_x}
%\end{equation}
where the second equation gives the relation between $x$ and a
macroscopic velocity component $V$ measured along the photon
propagation (i.e.\ parallel to the light path and in the same
direction).  $b$ is the usual Doppler parameter, $b = \sqrt{V_{th}^2 +
  V_{turb}^2}$.  Eq.\ \ref{eq_tau} shows that \lya\ is very rapidly
optically thick at line center, i.e.\ already for modest column
densities ($N_H > 3\times 10^{13}$ cm$^{-2}$).  For $N_H=10^{20}$ a
very large number of scatterings ($\sim 10^7$) are required to
escape. However, velocity fields or a inhomogeneous medium can ease
the escape (cf.\ below).  

\begin{figure}[tb]
%\centerline{\psfig{file=5554fg15.eps,width=6cm}
%  \psfig{file=5554fg17.eps,width=6cm}}
\centerline{\psfig{file=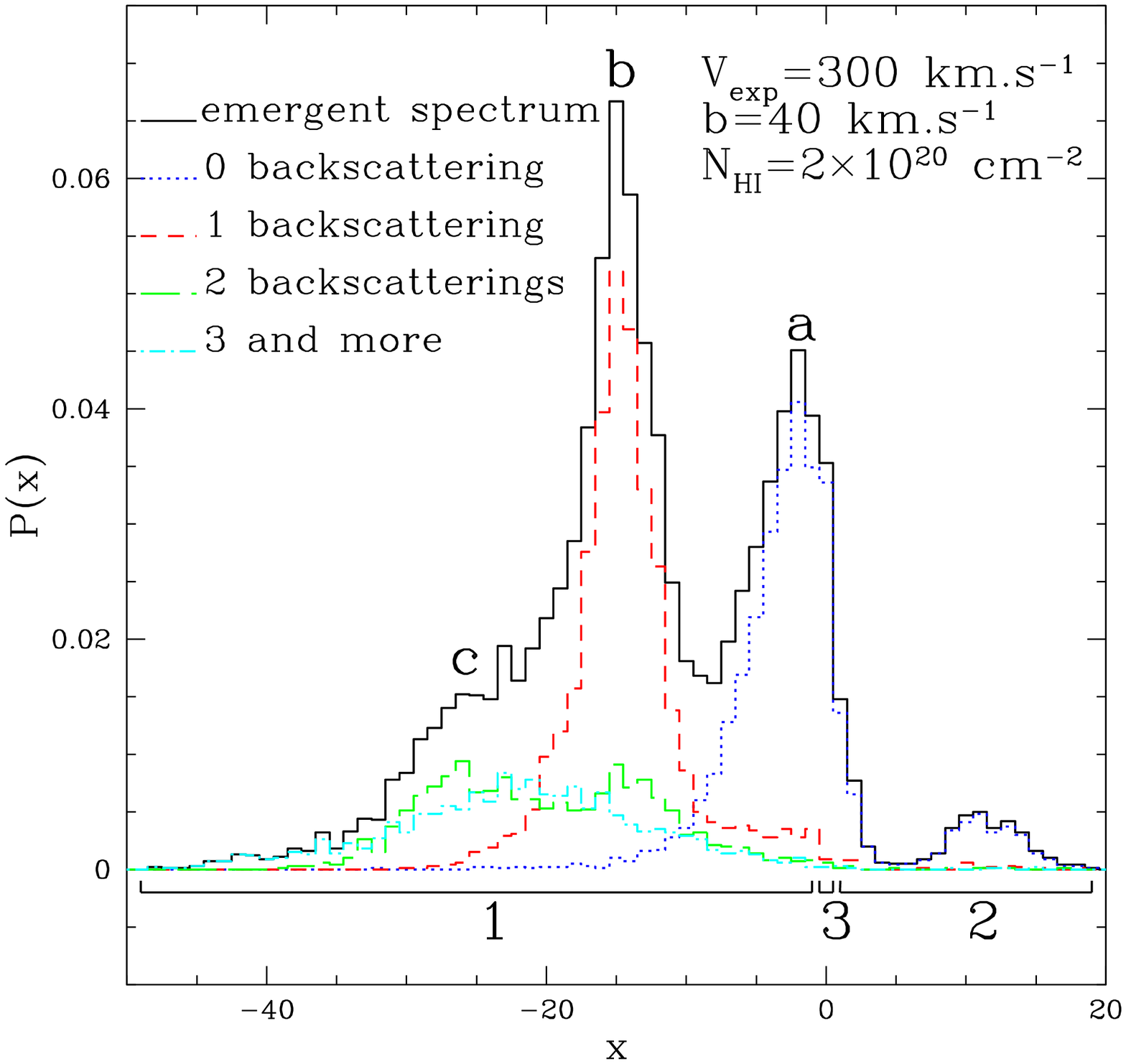,width=6cm}
  \psfig{file=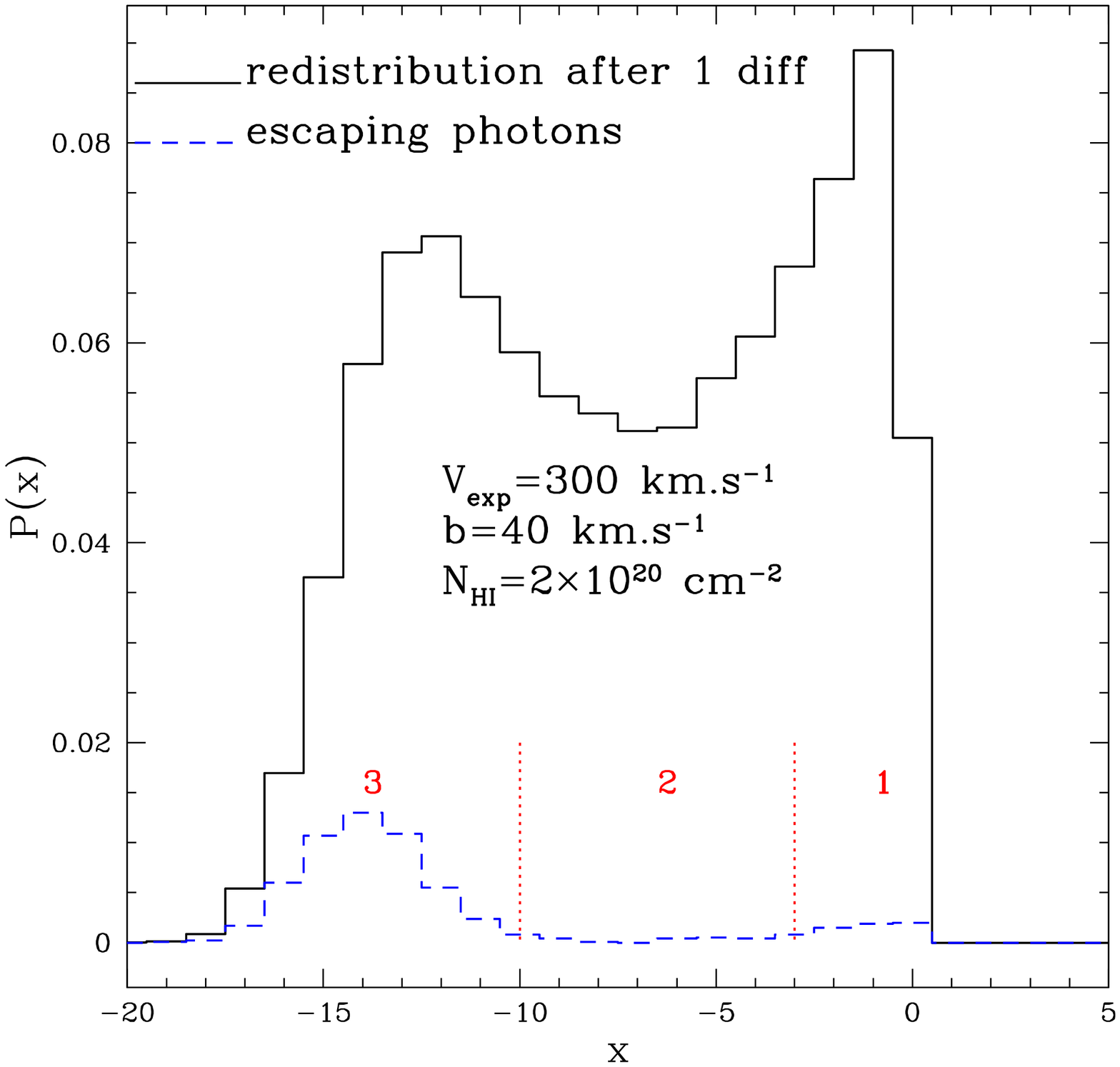,width=6cm}}
\caption{{\bf Left:} Emergent \lya\ profile from an expanding shell
with central monochromatic source. The different
  shapes can be described with the number of back-scatterings that photons
  undergo: bumps $1a$ and $2$ are built-up with photons that did not
  undergo any backscattering, the highest peak located at $x= -2
  \vexp/b$ (feature $1b$) is composed of photons that undergo
  exactly one backscattering, and the red tail $1c$ is made of photons
  that undergo two or more backscatterings. See Verhamme \etal\ (2006) 
  for more details.
{\bf Right:} Frequency distribution of the photons in the expanding shell
  after the first scattering. The black solid curve contains all
  photons, and the blue dotted one represents the histogram of photons
  which escaped after only one scattering.  
  They form a bump around $x \sim -2x(\vexp)$, which explains the
  appearance of feature 1b. See description in text. 
From Verhamme \etal\ (2006).}
\label{fig_shell}
\end{figure}

As true for other lines, the scattering of
photons in the \lya\ line is not a random walk: it corresponds to a
walk in coupled spatial and frequency space, where transport is
dominated by excursions to the line wings.  In other words, photons
propagate only over large distances allowing (long mean free path)
them to escape when they are in the wings, where the opacity is lower.
This already suffices to understand the formation of double peak \lya\
line profiles in the case \lya\ emission surrounded (or covered) by a
static medium, as shown in Fig.\ \ref{fig_freq} (left): all photons
initially emitted at line centre (for illustration) are absorbed and
``redistributed'' to the wings, where they can escape. The higher the
total optical depth, the larger the separation of the two
peaks becomes.
Asymmetries between the two peaks are of course introduced
with shifts of the intrinsic emission frequency, or
-- equivalently -- with an approaching/receding medium.
These cases and many variations thereof are discussed in detail by
Neufeld (1990).

In contrast to other scattering processes, \lya\ scattering is neither
coherent nor isotropic. The frequency redistribution, e.g.\ described
by the angle averaged frequency redistribution functions $R_{II}$ of
Hummer (1962), is illustrated in Fig.\ \ref{fig_freq} (right).
Schematically, for input frequencies $x_{in}$ close to the core the
emergent photon has its frequency redistributed over the interval
$\sim [-x_{\rm in},+x_{\rm in}]$. Once photons are sufficiently far in
wing they are re-emitted close to their input frequency, i.e.\
scattering is close to coherent in the comoving frame.
This behaviour is fundamental to understand e.g.\ the formation of the
emergent line profile for expanding shells, illustrated in Fig.\
\ref{fig_shell}.  There detailed radiation transfer calculations show
that the peak of the asymmetric \lya\ profile is located approximately
at the frequency Doppler-shifted by twice the expansion velocity
resulting from photons from the backside of the shell (see Verhamme
\etal\ 2006). This mostly results from two facts: first
the reemission of the photons after their first scattering in the shell
peaks at a Dopplershift of $\sim 1 \times \vexp$ in the comoving
reference frame of the shell, since the original \lya\ photon emitted
at line center ($x=0$) is seen in wing by the material in the shell
(reemission close to coherence). In the external frame these 
photons have then frequencies between $x \sim 0$ and $-2x(\vexp)$.
Now, the escape of the photons with the largest redshift being 
favoured, this will preferentially select photons from the back
of the shell, creating thus a peak at $-2x(\vexp)$. The interplay 
between these different probabilities imprints the detailed 
line shape, discussed in more detail in Verhamme \etal\ (2006).
For a given geometry, e.g.\ an expanding shell appropriate to model
outflows in starbursts, a wide variety of \lya\ profiles
can be obtained depending on the shell velocity and its temperature, 
the column density, on the relative strength of the initial
\lya\ emission with respect to the continuum, and on the presence
of dust (see Verhamme \etal\ 2006 for an overview).
Let us now briefly discuss how dust affects the \lya\ radiation
transfer.

\subsubsection{\lya\ transfer with dust}
A simple comparison of the probability of \lya\ photons to interact
with dust, 
$P_d= \frac{n_d \sigma_d}{n_H \sigma_{\rm{H}}(x) + n_{d} \sigma_d}$, 
shows that this event is quite unlikely, especially in
the line core, where the \lya\ cross section exceeds that of dust,
$\sigma_d$ by several orders of magnitudes. Despite this
interactions with dust particles occur, especially in the wings,
but also closer to line center since the overall
probability for a photon to interact with dust is increased by
the large number of line scattering occurring there.
For this reason it is immediately clear that the dust destruction
of \lya\ photons depends also on the kinematics of the \hi\
gas, where supposedly the dust is mixed in, although {\em per se}
the interaction of UV photons with dust is independent of the 
gas kinematics.

The net result is a fairly efficient destruction of \lya\ photons
by dust, as e.g.\ illustrated for static cases by Neufeld (1990),
and expanding shells by Verhamme \etal\ (2006). In the latter case
the escape of \lya\ photons is typically reduced by a factor
$\sim$ 2--4 with respect to a simple reduction by $\exp(-\tau_a)$,
where $\tau_a$ is the dust absorption optical depth. 
Finally it is also interesting to note that dust does not only reduce
the \lya\ emission (or the line equivalent width), it also alters
somewhat the line profile in a non-grey manner (cf.\ Ahn 2004, Hansen
\& Oh 2006), since its effect depends on \lya\ scattering. 
See Verhamme \etal\ (2006) for illustrations.

\subsubsection{\lya\ transfer: geometrical effects}
Given the scattering nature of \lya\ it is quite clear that the
observed \lya\ properties of galaxies depend also in particular on
geometry. By this we mean the intrinsic geometry of the object, i.e.\
the spatial location of the ``initial'' \lya\ emission in the \hii\
gas, the distribution and kinematics of the scattering medium (namely
the \hi), but also the spatial region of this object which is
ultimately observed. In other words the observed \lya\ line properties
(equivalent width and line profile) will in principle also vary if the
observations provide a integrated spectrum of the entire galaxy or
only a region thereof.

In an inhomogeneous ISM, UV continuum and \lya\ line photons
will also propagate in different ways, since their
transmission/reflection properties differ. Such cases
were e.g.\ discussed by Neufeld (1991) and Hansen \& Oh (2006),
who show that this can lead to higher \lya\ equivalent widths.

In non-spherical cases, including for example galaxies with strong
outflows and galactic winds with complex geometries and velocity
structures one may of course also expect significant orientation
effects on the observed \lya\ line.  Such cases remain largely to be
explored in realistic 3D radiation transfer simulations.

\subsection{Lessons from Lyman Break Galaxies}
Having already discussed relatively nearby starburst galaxies, where
spatial information is available, it is of interest to examine the
empirical findings related to \lya\ of more distant spatially
unresolved objects, the so-called Lyman Break Galaxies (LBG) discussed
also in more detail in Sect.\ \ref{s_dist} and by Giavalisco in this
volume.  These different categories of objects may help us
understanding in particular \lya\ emission and stellar populations in
distant and primeval galaxies.

\begin{figure}[tb]
%\centerline{\psfig{file=bestfit_4691.eps,width=4cm}
%  \psfig{file=bestfit_4454.eps,width=4cm}
%  \psfig{file=cB58.eps,width=4cm}}
\centerline{\psfig{file=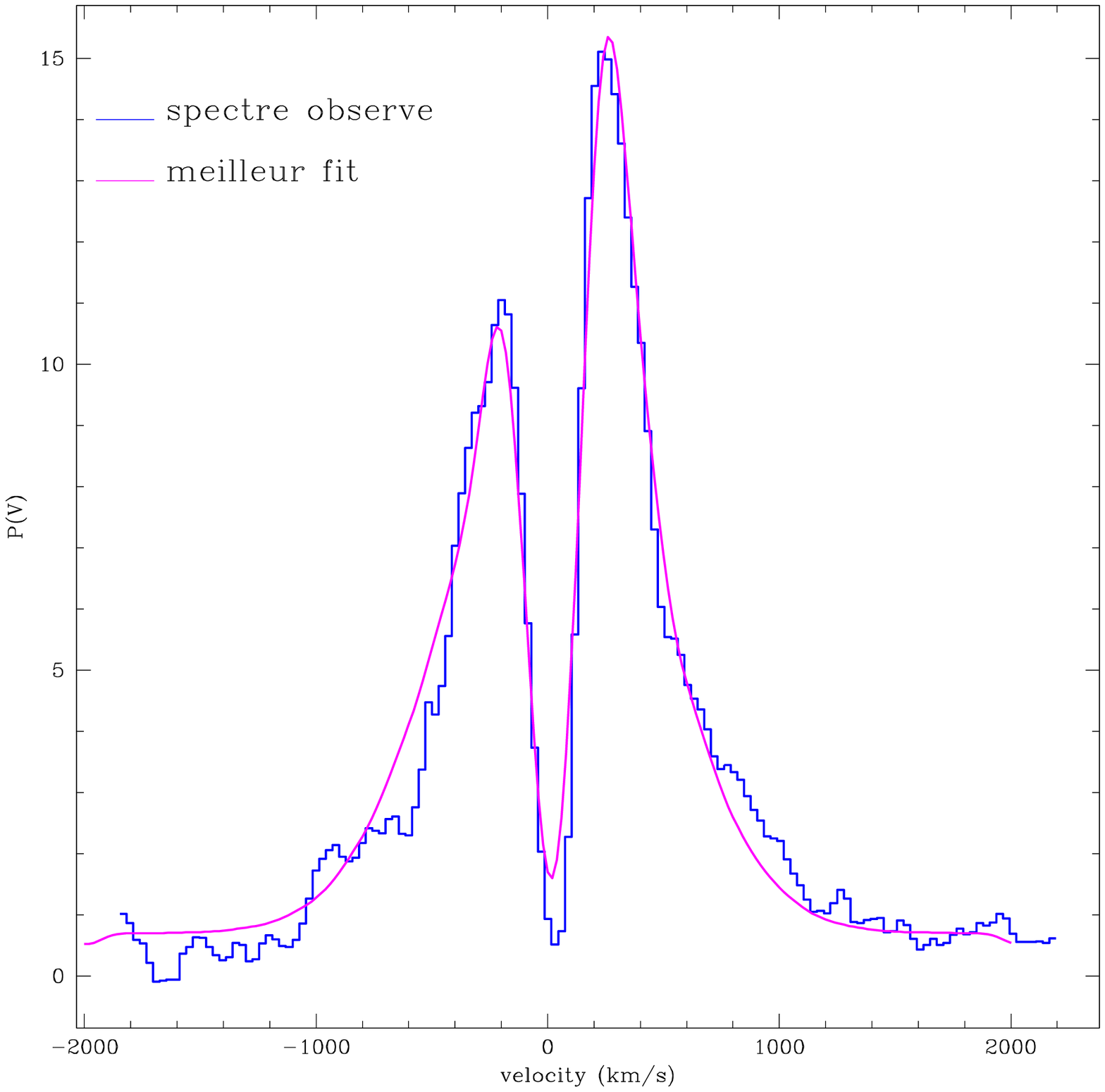,width=4cm}
  \psfig{file=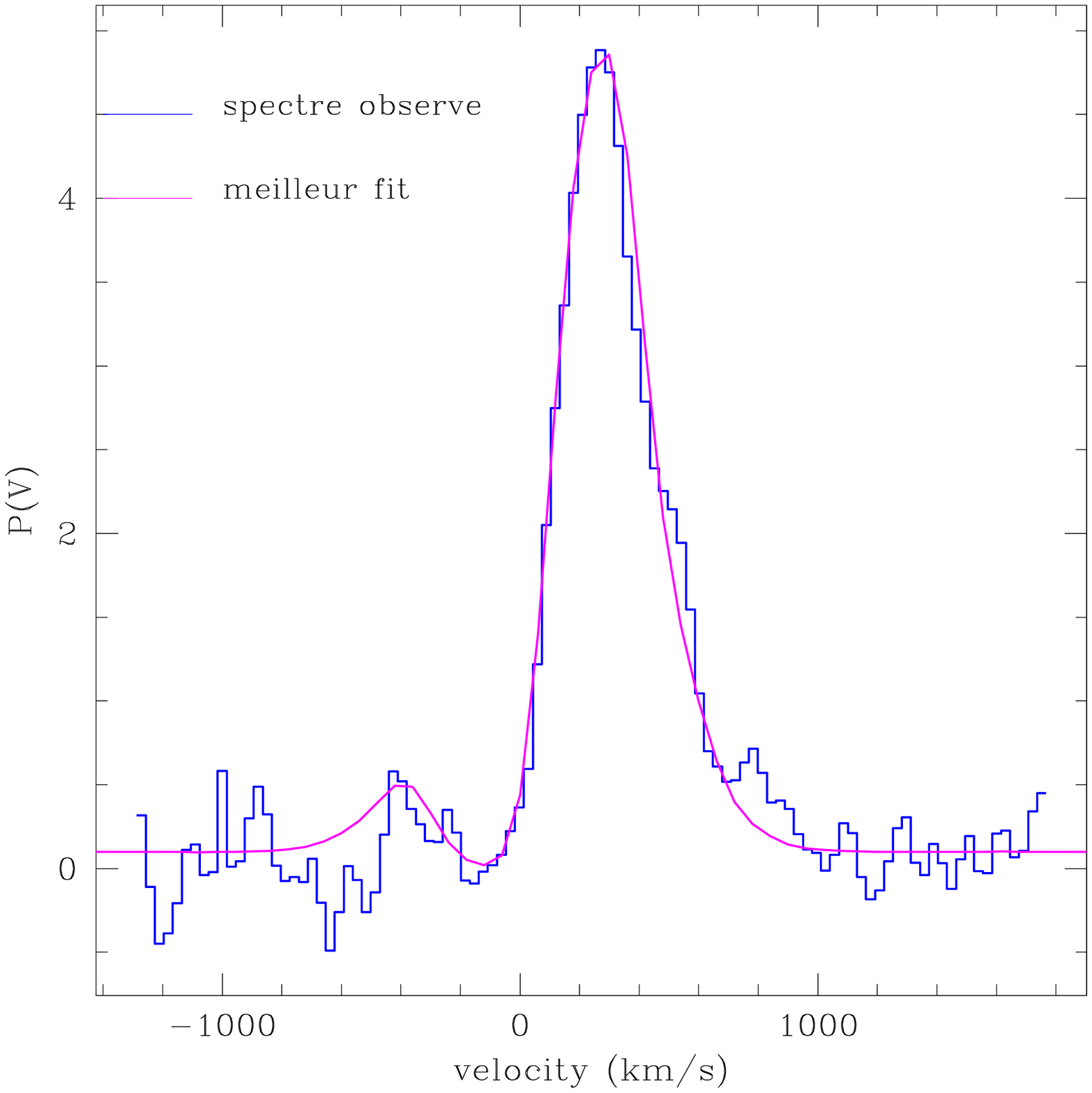,width=4cm}
  \psfig{file=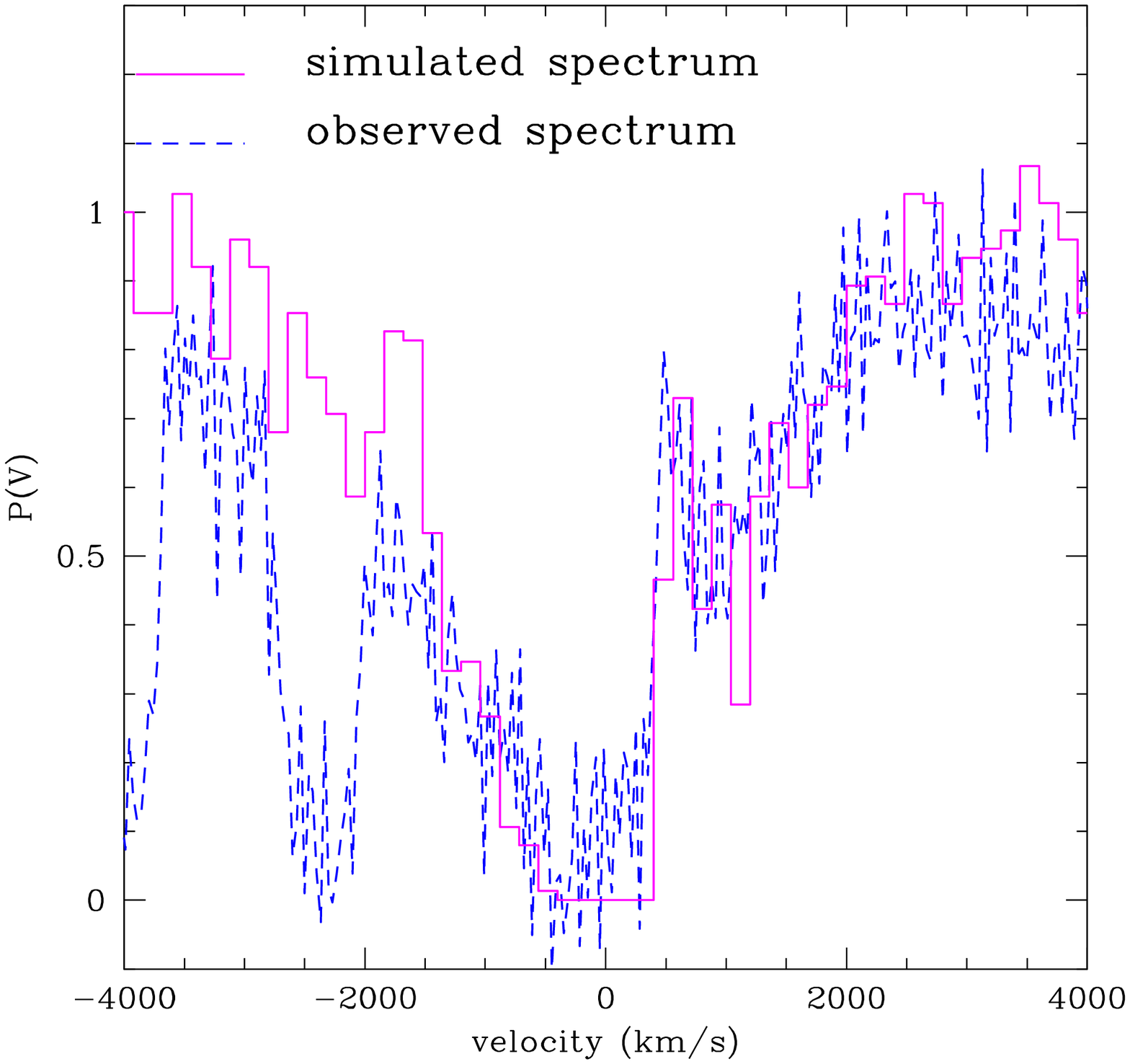,width=4cm}}
\caption{Comparison of observed and modeled \lya\ line profiles
of $z\sim 3$ LBGs showing a variety of different line profile 
morphologies, from double peaked, over P-Cygni, to broad absorption.
See discussion in text. From Verhamme \etal\ (2007).}
\label{fig_fdf}
\end{figure}	

LBGs are galaxies with intense ongoing star formation, selected from their UV
(restframe) emission. In 2003 approximately 1000 LBGs with
spectroscopic redshifts were know, mostly studied by the group of
Steidel (see Shapley \etal\ 2003).
Since then the number has grown, but this study remains 
the most comprehensive one on $z \sim 3$ LBGs.
The restframe UV spectra of LBGs show stellar, interstellar and nebular 
lines testifying of the presence of massive stars.
A diversity of \lya\ line profiles, ranging from emission, over P-Cygni to 
broad absorption line profiles, and different strengths are observed.
Interstellar (IS) lines are found blueshifted with respect to the
stellar lines (defining the object redshift, when detected) by 
$\Delta v({\rm abs}-\star)=-150 \pm 60$ \kms.  A shift of 
$\Delta v({\rm em-abs}) \sim$ 450--650 \kms\ is also observed between 
the IS absorption lines and \lya.
Finally Shapley \etal\ (2003) find several correlations between the
extinction, $W(\lya)$, $W(IS)$, and the star formation rate (SFR), which
are not or poorly understood, at least until very recently
(see Ferrara \& Ricotti 2007 for a possible explanation).

From \lya\ radiation transfer modeling discussed before, the observed
shifts between stellar, IS lines and \lya\ are naturally understood 
if the geometry is that of a ``global'' expanding shell
(Verhamme \etal\ 2006).
The IS lines are then formed by absorption of the UV continuum light
from a central starburst in the shell along the line of sight towards
the observer. Their bluedshift with respect to the stars measures
thus the expansion velocity \vexp. One then obtains naturally
$\Delta v({\rm em-abs}) \sim 3 \times |\Delta v({\rm abs}-\star)|
= 3 \vexp$, since \lya\ originates from the back of shell redshifted 
by $2 \vexp$.
This result indicates that large-scale, fairly symmetric shell
structures must be a good description of the outflows in LBGs.

What causes the variety of observed \lya\ line profiles and what does
this tell us about these galaxies?  Using the radiation transfer code
described in Verhamme \etal\ (2006) we have recently undertaken the first
detailed modeling of typical LBGs at $z \sim 3$, in particular objects
from the FORS Deep Field observed by Tapken \etal\ (2007) at a spectral
resolution $R \sim 2000$, sufficient to do detailed line profile
fitting. Assuming the spherically expanding shell model motivated 
in particular by the correct velocity shifts just mentioned,
the full variety of profiles can be reproduced for the observed values
of \vexp\ and extinction, and by varying \nh, and intrinsic
\lya\ line parameters ($W$ and FWHM).

Three such examples are illustrated in Fig.\ \ref{fig_fdf}.  Fitting
the double peak profile of FDF 4691 (left), is only possible with low
velocities, i.e.\ conditions close to a static medium (cf.\ Fig.\
\ref{fig_freq}). Such \lya\ profiles are relatively rare;
other cases with such double peak profiles include the \lya\
blob observed by Wilman \etal\ (2005) and interpreted by them as
a ``stalled'' expanding shell, or even as a collapsing protogalaxy
(Dijkstra \etal\ 2006b).
The profile of FDF 4454 (middle), quite typical of LBGs, indicates a
typical expansion velocity of $\vexp \sim$ 220 \kms\ and a low
extinction, compatible with its very blue UV slope.  Finally, the
profile of the lensed galaxy cB58 (right) from Pettini \etal\ (2000)
is well reproduced with the observed expansion velocity and extinction
($\vexp \sim$ 255 \kms, $E_{B-V}=0.3$).
The fits yield in particular constraints on the column density \nh\ 
and the intrinsic \lya\ line parameters ($W$ and FWHM).
This allows us to examine the use of \lya\ as a SFR indicator,
to provide constraints on the SF history and age of these galaxies,
and to shed new light on the observed correlations between 
\lya\ and other properties of LBGs (see Verhamme \etal\ 2007).
Understanding \lya\ in galaxies for which sufficient observations
are available and located at different redshift is of great
interest also to learn how to exploit the more limited information
available for objects at higher $z$, including primeval galaxies
(see Section \ref{s_dist}).

\subsection{\lya\  trough the InterGalactic Medium}
\label{s_lyaigm}
Having discussed the properties of \lya\ line formation and radiation
transfer effects in galaxies, we will now examine how
the \lya\ profile is transformed/transmitted on its way to the observer,
i.e.\ through the intergalactic medium (IGM).

In this situation we consider radiation from a distant
background source passing trough one or several ``\hi\ clouds''.
This geometry leads to a very simple case where \lya\
photons are absorbed and then either scattered out of the line of
sight or absorbed internally by dust. In other words {\em no
true radiation transfer needs to be computed}, and the
resulting \lya\ profile of the radiation emerging from the cloud
is simply the input flux attenuated by a Voigt absorption profile
characteristic of the cloud properties. 
For a given density and (radial) velocity -- or equivalently redshift --
distribution along the line of sight, the computation of the
total attenuation and hence of the observed spectrum is thus
straightforward.

The observational consequences for a distant source will thus be:
{\em 1)} the imprint of a number of (discrete) absorption components
on top of the background source spectrum due to intervening \hi\ 
clouds or filaments, and
{\em 2)} an alteration of the emergent galactic \lya\ profile 
plus a reduction of the \lya\ flux if neutral H is present close
in velocity/redshift space to the source.
The first is well known observationally as the 
\lya\ forest, leading even to a complete absorption (the so-called 
Gunn-Peterson trough) in distant ($z \sim 6$) quasars
(see the review by Fan \etal\ 2006).
The appearance of a complete Gunn-Peterson trough in high-$z$ quasars
implies a quantitative change of the ionisation of the IGM,
possibly tracing the end of the epoch of cosmic reionisation
(cf.\ Fan \etal\ 2006).
The second effect leads e.g.\ to the alteration of the \lya\ profile
and to a strong reduction of the \lya\ flux in high-$z$ quasar,
due to absorption by the red damping wing of \lya\ by nearby
\hi\ (cf.\ Miralda-Escud\'e 1998, and observations by Fan \etal\ 2003).

The two effects just discussed have the following immediate implications:
\begin{itemize}
\item The SED of high-$z$ galaxies is altered by Lyman-forest attenuation
at wavelengths shorter than \lya\ ($< 1216$ \AA). A statistical
description of this attenuation is given by Madau (1995).
\item For $z \ga$ 4--5 the Lyman-forest attenuation is so strong that
it effectively leads to a spectral break at \lya, replacing therefore
the ``classical'' Lyman-break (at 912 \AA) due to photoelectric absorption
by H~{\sc i}). The \lya-break becomes then the determining feature
for photometric redshift estimates.
\item The reduction of the \lya\ flux implies that
{\em a)}  determinations
of the SFR from this line will underestimate the true SFR,
{\em b)} the observed \lya-luminosity function (LF) does not correspond
to the true (intrinsic) one, and
{\em c)} the detectability of high-$z$ \lya\ emitters (hereafter LAE)
is reduced.
%and one may indeed wonder if and how LAE are
%observable beyond
\item The \lya\ profile, \lya\ transmission, and the \lya\ luminosity
function contain information on the ionisation fraction of hydrogen
and can hence in principle constrain cosmic reionisation.
\end{itemize}

We will now discuss how/if it is still possible to observe LAE
beyond the reionisation redshift.
%, and how the LF(\lya) is 
%used to constrain cosmic reionisation.

\begin{figure}[tb]
%\centerline{\psfig{file=lya_priorreionis.eps,width=12cm}}
%\centerline{\psfig{file=figure13schaerer.eps,width=12cm}}
\centerline{\psfig{file=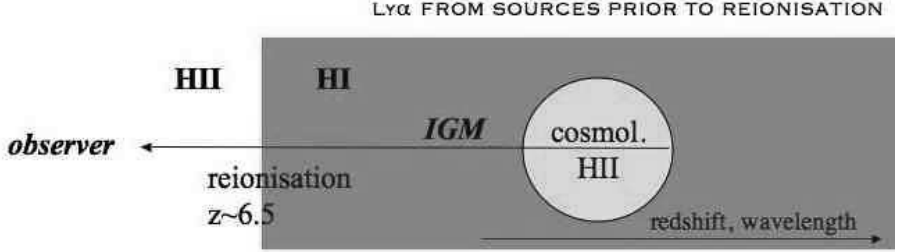,angle=270,width=12cm}}
\caption{Schematic representation of a star forming galaxy
situated beyond the reionisation redshift (here indicated at
$z_r \sim 6.5$), its surrounding cosmological \hii\ region,
the neutral IGM down to $z_r$, and the transparent (ionised)
IGM towards the observer. Redshift and the observed 
\lya\ wavelength increase to the right.}
\label{fig_reion}
\end{figure}	

\subsection{\lya\  from sources prior to reionisation}
How is it possible to observe \lya\ emission from sources ``beyond the
end of reionisation'', i.e.\ at very high redshift where the IGM
contains a significant fraction of neutral hydrogen which absorbs the
\lya\ emission?  The way to achieve this is in principle quite simple
and sketched in Fig.\ \ref{fig_reion}. It suffices to create around
the \lya\ source a ``cosmological'' \hii\ region big enough 
so that no or very little \hi\ is present at velocities -- i.e.\
redshifts -- close to the source. In this way the attenuation 
close to the \lya\ emission is avoided and the 
line flux from this distant source can propagate 
freely to the observer, since it comes from the most redshifted
part along the line of sight.

So, how are these cosmological \hii\ regions created? Obviously this
requires one or several sources (galaxies or quasars) producing
ionising photons which are able to escape the galaxy and can then
progressively ionise the surrounding IGM. This is referred to as the
``proximity effect''.
The properties and the evolution of cosmological \hii\ regions
have been studied and described analytically in several papers
(see e.g.\ Shapiro \& Giroux 1987, Cen \& Haiman 2000, and review
by Barkana \& Loeb 2001).
For example, neglecting recombinations in the IGM (since 
for the low IGM densities the recombination timescale is much longer
than the Hubble time) and assuming that the ionising source 
is ``turned on'' and constant during the time $t_Q$ the 
Stroemgren radius (size) of the \hii\ region becomes
\begin{equation}
R_{t_Q}=\left[ \frac{3 \dot{N}_{ph} t_Q}{4 \pi <n_H>}\right]^{1/3},
\end{equation}
where $\dot{N}_{ph} = f_{esc} Q_H$ is escaping ionising flux and
$<n_H>$ the mean IGM density taking possibly a non-uniform density
distribution into account.  The residual \hi\ fraction inside the
\hii\ region is given by photoionisation equilibrium and can also be
computed. Then the resulting attenuation $e^{-\tau}$ can
be computed by integrating the optical depth along the line
of sight
\begin{equation}
\tau(\lambda_{obs},z_s)= \int_{z_r}^{z_s} dz c \frac{dt}{dz} n_H(z) 
                         \sigma_\alpha(\lambda_{obs}/(1+z)).
\label{eq}
\end{equation}
Here $z_s$ is the source redshift, $z_r$ a limiting redshift (the redshift
of reionisation in Fig.\ \ref{fig_reion}) below which the IGM is
supposed to be transparent, and $\sigma_\alpha$ is the \lya\
absorption cross section. 

\begin{figure}[tb]
%\centerline{\psfig{file=haiman_2002_fig1.eps,width=6cm}
%  \psfig{file=haiman_2002_fig2.eps,width=6cm}}
%\centerline{\psfig{file=figure14schaerer.eps,width=6cm}
%  \psfig{file=figure15schaerer.eps,width=6cm}}
\centerline{\psfig{file=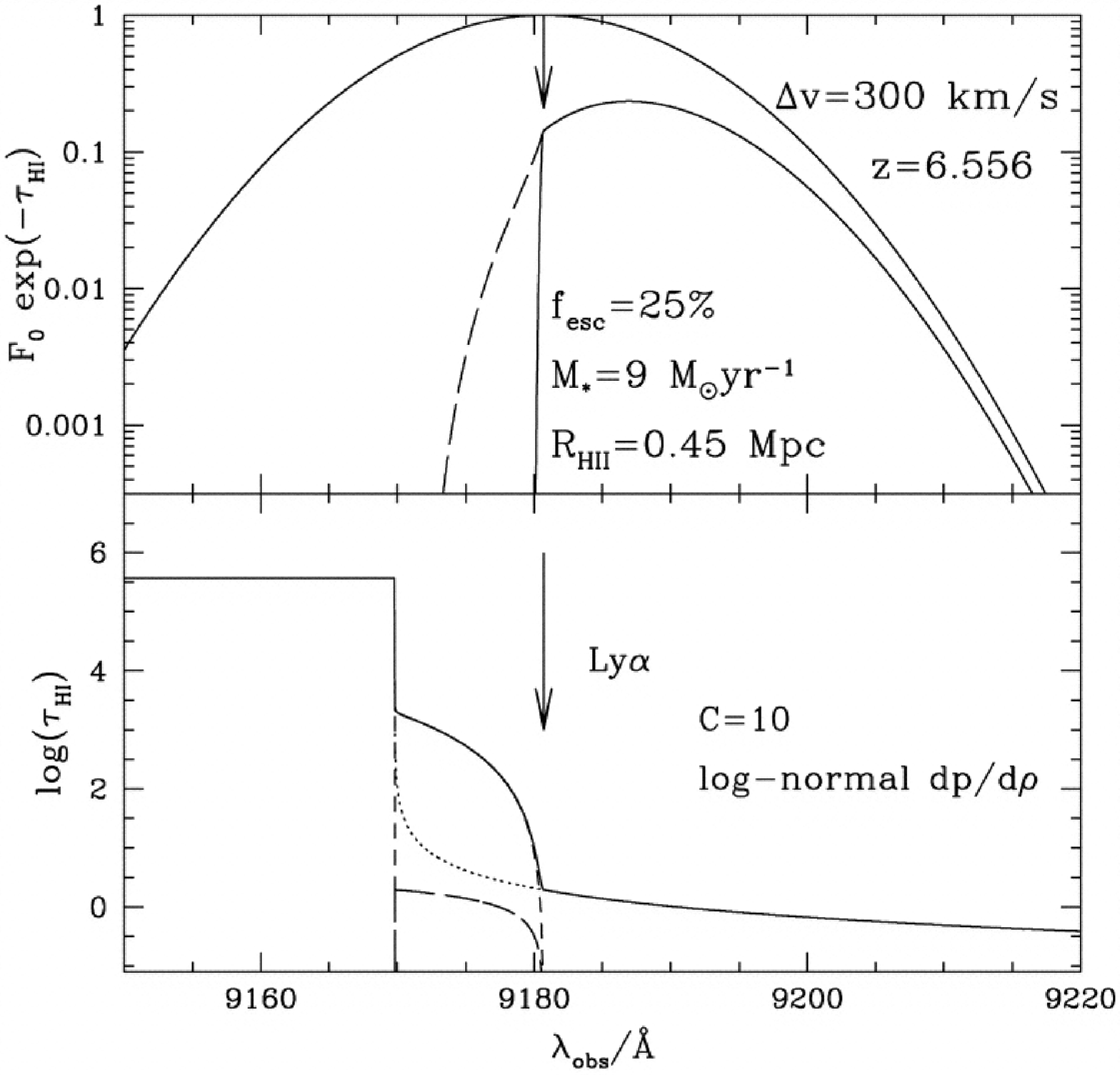,width=6cm}
  \psfig{file=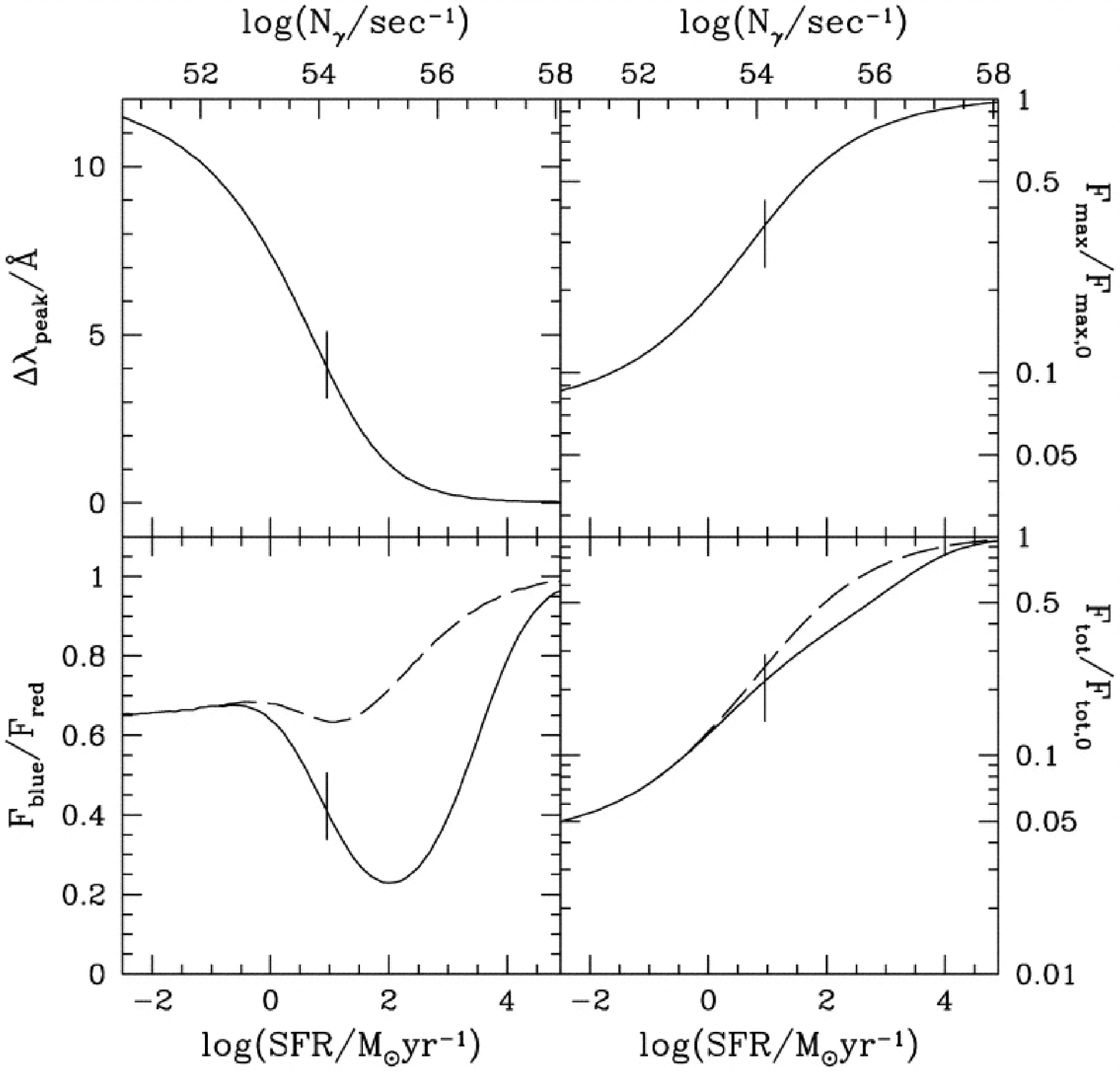,width=6cm}}
\caption{Predicted \lya\ line profile, \lya\ transmission and other
properties from the model of a $z=6.56$ lensed galaxy
taking IGM absorption into account. from Haiman (2002).
{\bf Left:} Intrinsic and resulting line profile (top), opacities
leading to the \lya\ attenuation.
{\bf Right:} Parameters, such as asymmetry, peak position, and total transmission
(bottom right) of the predicted \lya\ line as a function of the SFR.}
\label{fig_haiman}
\end{figure}	
For example, the observability of \lya\ from a $z=6.56$ 
galaxy observed by Hu \etal\ (2002) has been examined with
such a model by Haiman (2002). The results are illustrated
in Fig.\ \ref{fig_haiman}.
For a source with $SFR = 9$ \msunyr, an age of $\sim$ 100 Myr,
and an escape fraction $f_{esc} = 25$ \% the proper (comoving)  radius
of the \hii\ region is approximately 0.45 (3) Mpc. 
Assuming an intrinsic Lya profile with a width of $FWHM=300$ \kms\
Haiman obtains a transmission of $\sim$ 16\% of the \lya\ flux and
an asymmetric line profile, as observed.
A wider range of transmission encompassing also this value
is found from an independent estimate based on stellar 
population modeling (cf.\ Schaerer \& Pell\'o 2005).

In the picture described above, the \lya\ transmission is expected
to increase with increasing $SFR$, escape fraction, source lifetime,
and intrinsic line width, as also shown in Fig.\ \ref{fig_haiman} (right). 
The first three increase the size
of the cosmological \hii\ region; with the latter a higher fraction
of the line flux is emitted far from line center reducing thus
the absorption by the red damping wing in the \hi.
Other factors also affect the \lya\ transmission and the resulting
line profile: IGM infall, outflows (galactic winds), peculiar velocities
of the emitting gas within halo, the halo mass etc.
See Haiman (2002) and Santos (2004), which have examined these effects.

In a more realistic setting several ``complications'' can occur to
this simple model
(see e.g.\ Gnedin \& Prada 2004, Furlanetto \etal\ 2004, Wyithe \& Loeb 2004).
\begin{itemize}
\item  Clustering of sources helps to create a
larger \hii\ region.  Since the clustering probability increases with
$z$ and for fainter galaxies, this could play an important role 
for the detectability of high redshift \lya\ sources.  
\item In a non-homogeneous structure around the source the \hii\
regions are expected to deviate from spherical symmetry, since the
ionisation fronts will propagate more rapidly into directions with a
lower IGM density. 
\end{itemize}
From this it is clear that strong variations depending on the object,
its surroundings, and the viewing direction are expected and the
simple scaling properties of the spherical models described before may
not apply. A statistical approach using hydrodynamic simulations will
be needed.

In short, the answer to the question ``Is \lya\ emission from sources
prior to reionisation detectable?'' is affirmative from the theoretical
point of view, but the transmission depends on many factors! In any case,
searches for such objects are ongoing (cf.\ Sect.\ \ref{s_lae})
and will provide the definite answer.

\subsection{\lya\  Luminosity Function and reionisation}
As a last illustration of the use of \lya\ in distant, primeval 
galaxies we shall now briefly discuss the statistics of LAE,
in particular the \lya\ luminosity function LF(\lya), how it may be
used to infer the ionisation fraction of the IGM at different 
redshift, and difficulties affecting such approaches.

\begin{figure}[tb]
%\centerline{
  \begin{minipage}{0.45\linewidth}
    \psfig{file=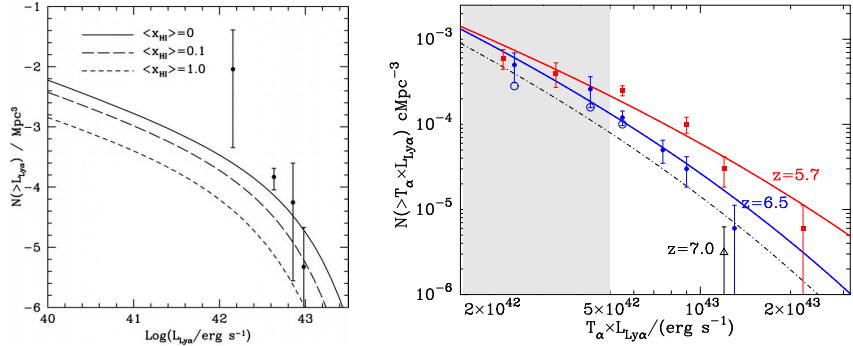,width=4.5cm}
  \end{minipage}
  \begin{minipage}{0.45\linewidth}
	\psfig{file=figure17schaerer.eps,width=4.5cm,angle=270}
  \end{minipage}
\caption{{\bf Left:} predicted \lya\ LFs for a fully ionised IGM (no attenuation case, 
i.e.\, the $z = 5.7$ LF; solid curve), and for an IGM with an increasing neutral
H fraction $x_{HI}$. From Haiman \& Cen (2005).  
{\bf Right:} Predicted and observed \lya\ LF at $z=5.7$ ad 6.5. 
The LF model is that by Dijkstra \etal\ (2006c). According to these authors
the observed ecline of the \lya\ LF 
attributed to the evolution of the halo mass function hosting the \lya\ emitters.}
\label{fig_lyalf}
\end{figure}

Since, as discussed above, the presence of neutral hydrogen in the IGM
can reduce the \lya\ flux of galaxies, it is clear that the \lya\ LF
is sensitive to the ionisation fraction $x_{HI}$. If we knew the intrinsic
LF$(z)$ of galaxies at each redshift, a deviation of the observed
LF from this intrinsic distribution could be attributed to attenuation
by \hi, and hence be used to infer $x_{HI}$
(cf.\ Fig.\ \ref{fig_lyalf}).
In practice the approach is of course to proceed to a differential comparison
of LF(\lya) with redshift.
Indeed, from simple \lya\ attenuation models like the ones
described in the previous section, a rapid decline of the LF is expected
when approaching the end of reionisation.

Haiman \& Spaans (1999) were among the first to advocate the use of LF(\lya)
and to make model predictions.
Since then, and after the detection of numerous LAEs allowing the measurement
of the \lya\ LF out to redshift $z=6.5$ (cf.\ Section
\ref{s_lae}), several groups have made new predictions of the \lya\ LF
and have used it to constrain cosmic reionisation. Some prominent
examples are Malhotra \& Rhoads (2004), Le Delliou \etal\ (2005, 2006), 
and Furlanetto \etal\ (2006).

One of the most recent of such attempts is presented by Dijkstra
\etal\ (2006c) who predict the \lya\ LF based on a modified
Press-Schechter formalism and introducing two main free parameters, a
star formation duty-cycle $\epsilon_{DC}$ and another parameter
depending on the SF efficiency, the escape fraction and the \lya\
transmission of the IGM.  They find a typical IGM transmission 
of $T_\alpha \sim$ 30 \% at $z=5.7$.
Adjusting the observed LFs at $z=5.7$ and
6.5 (where Quasars already indicate a significant change of the
ionisation fraction $x_{HI}$ as discussed in Sect.\ \ref{s_lyaigm})
Dijkstra \etal\ (2006c) find good fits without the need for a strong
change of the ionisation state advocated in other studies
(see Fig.\ \ref{fig_lyalf})  The
observed decline of the \lya\ LF between $z=5.7$
and 6 is attributed to the evolution of the halo mass
function hosting the \lya\ emitters. In this case this may
translate to a lower limit of $\sim$ 80\% for the fraction of ionised H
at $z=6.5$. This serves to illustrate the potential of LF(\lya)
analysis, but also the potential difficulties and the room for 
improvements.

Finally let us also note that Hu \etal\ (2005) do not find an evolution
of the mean \lya\ line profile between $z=5.7$ and 6.5, in agreement
with the above conclusion.

%%%%%%%%%%%%%%%%%%%%%%%%%%%%%%%%%%%%%%%%%%%%%%%%%%%%%%%%%%%%%%%%%%%%%%%%
%\section{Distant/primeval galaxies: searches and main results}
\section{Distant/primeval galaxies: observations and main results}
\label{s_dist}

Before we discuss searches for distant galaxies, provide an
overview of the main results, and discuss briefly open questions
we shall summarise the basic observational techniques used
to identify high redshift galaxies.

\subsection{Search methods}
The main search techniques for high-z galaxies can be classified 
in the two following categories.
\begin{enumerate}
\item The Lyman break or drop-out technique, which selects galaxies
over a certain redshift interval by measuring the Lyman break,
which is the drop of the galaxy flux in the Lyman continuum
(at $\lambda < 912$ \AA) of the \lya\ break (shortward of \lya)
for $z \ga$4--5 galaxies (cf.\ above).
This method requires the detection of the galaxy in several
(sometimes only 2, but generally more) broad-band filters.
\item Emission line searches (targeting \lya\ or other emission lines).
Basically three different techniques may be used:
{\em 1)} Narrow Band (NB) imaging (2D) e.g.\ of a wide field selecting
a specific redshift interval with the transmission of the NB filter.
Long slit spectroscopy (1D) for ``blind searches'' e.g.\ along 
critical line in lensing clusters, or 
observations with Integral Field Units (3D) allowing to explore
all three spatial directions (2D imaging + redshift). 
The first one is currently the most used technique.
\end{enumerate}
In practice, and to increase the reliability, several methods
are often combined.

Surveys/searches are being carried out in blank fields or targeting
deliberately gravitational lensing clusters allowing one to benefit
from gravitational magnification from the foreground galaxy cluster.
For galaxies at $z \la 7$ the Lyman-break and \lya\ is found in the
optical domain. Near-IR ($\ga$ 1 \micron) observations are necessary 
to locate $z \ga 7$ galaxies. 
%In principle such observations
%could allow the detection of galaxies out to redshifts of $\sim$
%13--18 from the ground (in the K band). 
% ground-based, … JWST z >~13-18: JWST only !

The status in 1999 of search techniques for distant galaxies
has been summarised by Stern \& Spinrad (1999). For 
more details on searches and galaxy surveys see the lecture
notes of Giavalisco (these proceedings).

\subsection{Distant \lya\ emitters}
\label{s_lae}
Most of the distant known \lya\ emitters (LAE) have been found
through narrow-band imaging with the SUBARU telescope, thanks to 
its wide field imaging capabilities. $z \sim$ 6.5--6.6 LAE
candidates are e.g.\ selected combining the three
following criteria: an excess in a narrowband filter (NB921)
with respect to the continuum flux estimated from the broad
$z^\prime$ filter, a 5 $\sigma$ detection in this NB filter,
and an $i$-dropout criterium (e.g.\ $i-z^\prime>1.3$) making
sure that these objects show a \lya\-break. 
Until recently 58 such LAE candidates were found,
with 17 of them confirmed subsequently by spectroscopy
(Taniguchi \etal\ 2005, Kashikawa \etal\ 2006).
The Hawaii group has found approximately 14 LAE at $z \sim 6.5$
(Hu \etal\ 2005, Hu \& Cowie 2006). 
The current record-holder as the most distant galaxy
with a spectroscopically confirmed redshift of $z=6.96$
is by Iye \etal\ (2006).
Six candidate \lya\ emitters between $z=8.7$ and 10.2 were
recently proposed by Stark \etal\ (2007) using 
blind long-slit observations along the critical lines in lensing
clusters.

LAE have for example been used with SUBARU to trace large scale structure 
at $z=5.7$ thanks to the large field of view (Ouchi \etal\ 2005).

Overall, quite little is known about the properties of NB selected
LAE, their nature and their relation to other galaxy types (LBG and
others, but cf.\ Sect.\ \ref{s_lbg}), since most of them --
especially the most distant ones -- are detected in very few bands,
i.e.\ their SEDs is poorly constrained.  The morphology of the
highest-$z$ LAEs is generally compact, indicating ionised gas with
spatial extension of $\sim$ 2--4 kpc or less (e.g.\ Taniguchi \etal\ 2005,
Pirzkal \etal\ 2006).

Although showing SF rates (SFR) of typically 2 to 50 \msunyr, 
the SFR density of LAE is only a fraction of that of LBGs at all redshifts.
%LAE show
%a lower space density and hence their SFR density is only a fraction
%of that of LBGs at all redshifts.  
For example at $z \sim$ 5--6.5, Taniguchi \etal\ (2005) estimate
the star formation rate density (SFRD) from \lya\ emitters as
SFRD(LAE) $\sim 0.01 \times$ SFRD(LBG), or up to 10 \% of SFRD(LBG) at
best if allowing for LF corrections.  At the highest $z$ this value
could be typically $3 \times$ higher if the IGM transmission of $\sim$
30\% estimated by Dijkstra \etal\ (2006c) applies.  Shimasaku \etal\
(2006) have found a similar space density or UV LF for LAE and LBG at
$z \sim 6$, and argue that LAEs contribute at least 30 \% of the SFR
density at this redshift.

The typical masses of LAE are still uncertain and being debated. For
example, Lai \etal\ (2007) find stellar masses of $M_\star \sim 10^9$
and $10^{10}$ \msun\ for three LAE at $z \sim 5.7$, whereas Prizkal
\etal\ (2006) find much lower values of $M_\star \sim 10^6$ and
$10^{8}$ \msun\ for their sample of $z \sim 5$ \lya\ galaxies.
Finkelstein \etal\ (2006) find masses between the two ranges for
$z \sim 4.5$ LAEs. Selection criteria may explain some of these
differences; e.g.\ the Lai \etal\ objects were selected
for their detection at 3.6 and 4.5 \micron\ with Spitzer.
Mao \etal\ (2006) argue that LAEs are limited to a relatively 
narrow mass range around $M_\star \sim 10^9$ \msun.
Further studies will be necessary to properly understand
the connections between LBG and LAE and the evolution of the two 
populations with redshift.

%Two key results:
%L emitters less significant than dropouts as contributors to SFR at z~6.6
%Yet an increasing fraction with increasing redshift (less evolution from z~3-6 than dropouts)

%also Overzier \etal\ (LAE at $z=4.1$) and
%Gawiser \etal\ (2006) at $z \sim 3$.

\subsubsection{PopIII signatures in LAE?}
\begin{figure}[tb]
%\centerline{\psfig{file=malhotra_rhoads2002_fig1.eps,width=6cm}
%  \psfig{file=MS2980f7a.eps,width=6cm}}
\centerline{\psfig{file=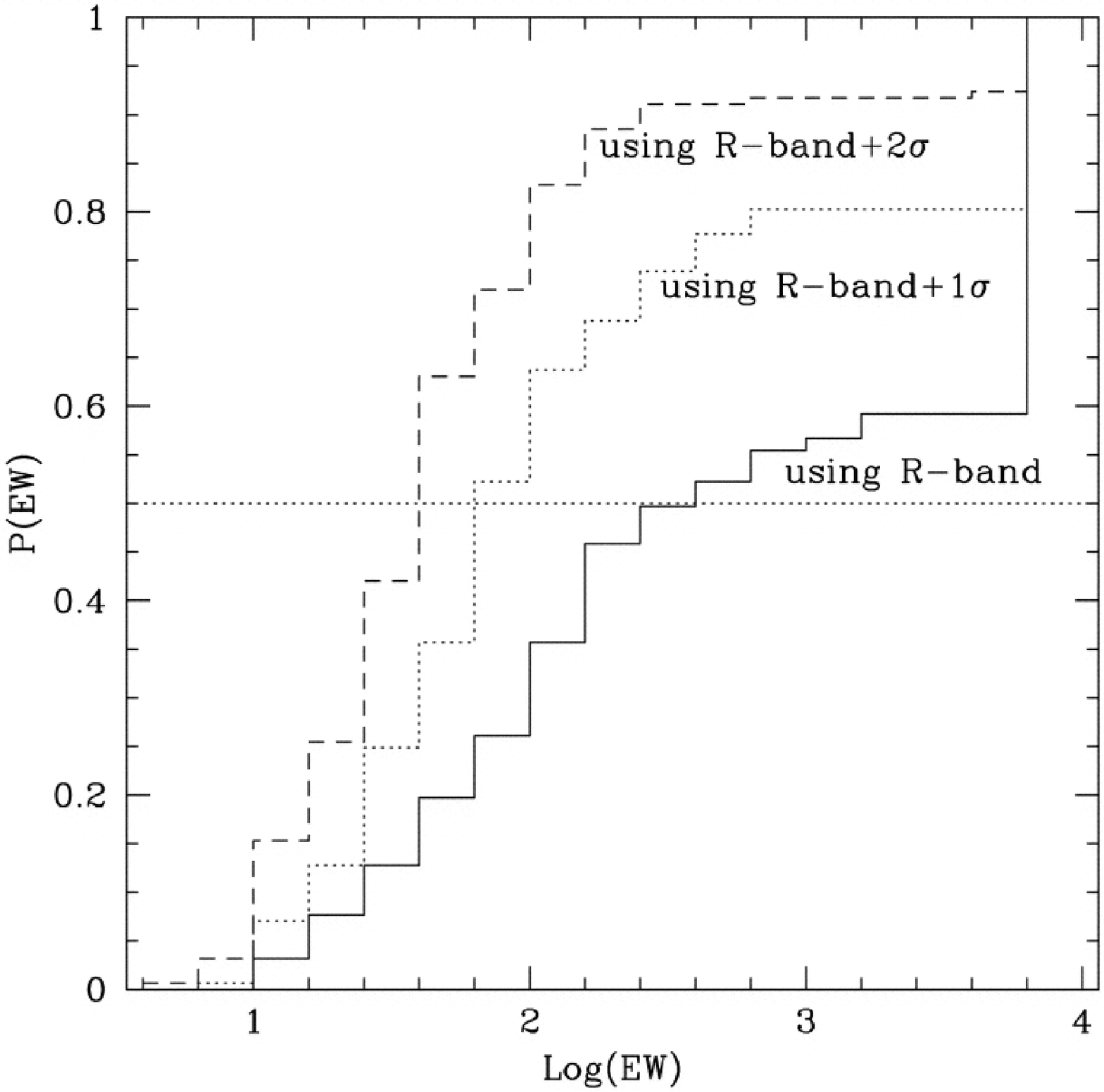,width=5.5cm}
%\centerline{\psfig{file=figure18schaerer.eps,width=5.5cm}
  \psfig{file=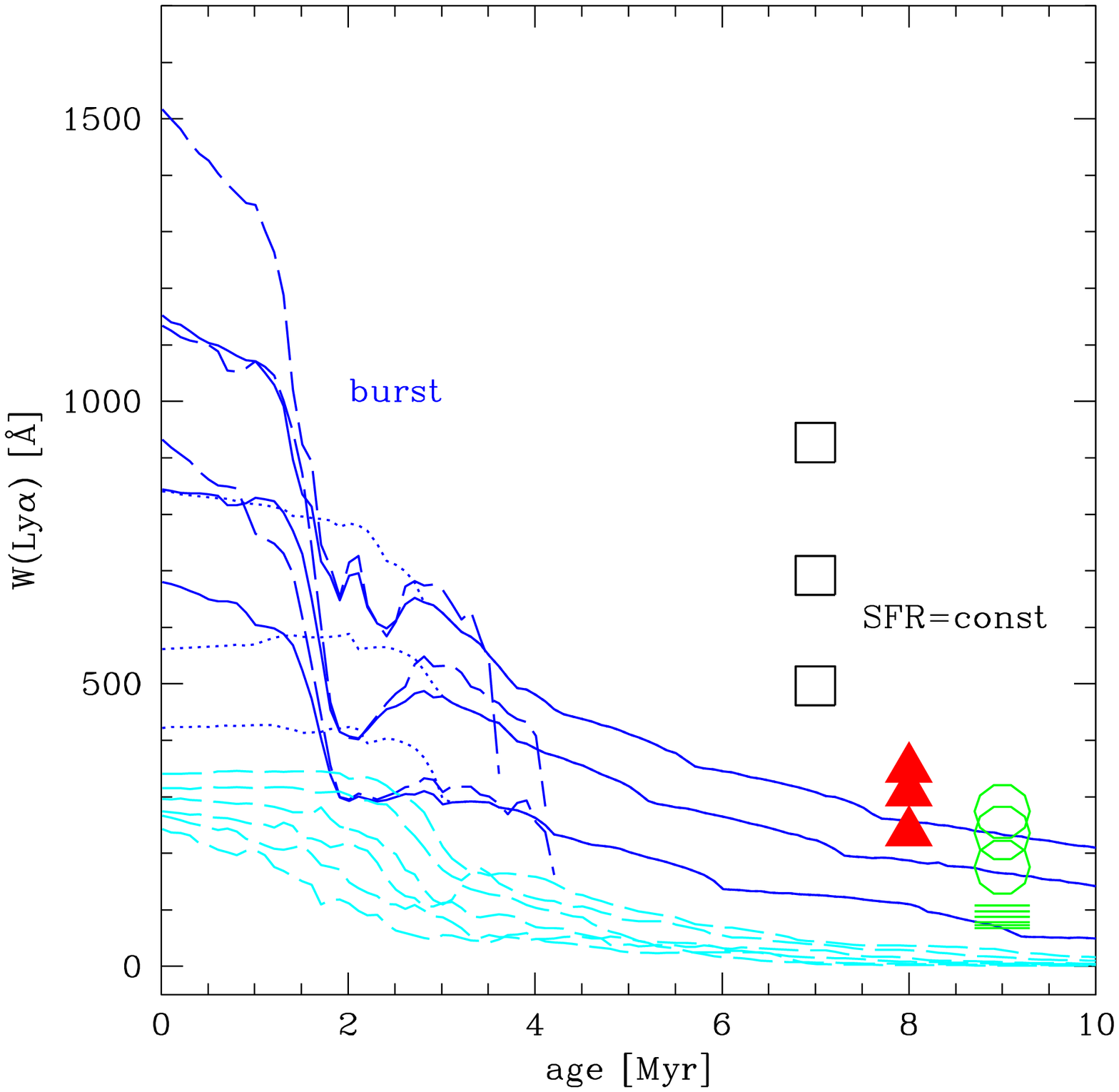,width=6cm}}
\caption{{\bf Left:} Observed \lya\ equivalent width distribution of 
$z=4.5$ sources from the LALA survey From Malhotra \& Rhoads (2002).
{\bf Right:}  Predicted \lya\ equivalent width for starbursts at different
metallicities (from solar to PopIII). Normal metallicities($Z \protect\ga 1/50$ \zsun) are 
shown by the magenta dashed lines. The maximum value predicted in 
this case is $W($\lya$ ) \sim 300$ \protect\AA. From Schaerer (2003). }
\label{fig_lala}
\end{figure}

The Large Area Lyman-$\alpha$ (LALA) survey by Rhoads and
collaborators, carried out on 4m class telescopes, has been one of the
first to find a significant number of LAE at high redshift ($z=4.5$,
5.7, and later also objects at 6.5). Among the most interesting
results found from LALA is the finding of a large fraction of LAE with
an apparently high median \lya\ equivalent width, compared to
expectations from normal stellar populations (see Fig.\
\ref{fig_lala}). Indeed, half of their $z=4.5$ candidates show
$W(\lya)$ in excess of $\sim$ 200--300 \AA\ (Malhotra \& Rhoads 2002),
a value expected only for very young stabursts, populations with
extreme IMFs, or very metal-poor (or PopIII) stars (cf.\ Schaerer
2003).  Malhotra \& Rhoads (2002) suggested that these could be AGN or
objects with peculiar top-heavy IMFs and/or PopIII dominated.  In this
context, and to explain other observations, Jimenez \& Haiman (2006)
also advocate a significant fraction of PopIII stars, even in $z \sim$
3--4 galaxies.
Recently Hansen \& Oh (2006), reviving an idea of
Neufeld (1991), have suggested that the observed $W(\lya)$ could be
``boosted'' by radiation transfer effects in a clumpy ISM.

Follow-up observations of the LALA sources have allowed to exclude the
narrow-line AGN ``option'' (Wang \etal\ 2004), but have failed to
provide further explanations of this puzzling behaviour. A fraction of
$\sim$ 70\% of the LALA LAE have been confirmed spectroscopically;
some high equivalent widths measurement could also be confirmed
spectroscopically \footnote{But aperture effects may still lead to an
  overestimate of $W(\lya)$.}.  Deep spectroscopy aimed at detecting
other emission lines, including the \Heiiuv\ line indicative of a
PopIII contribution (cf.\ Sect.\ \ref{s_obsprop}), have been
unsuccessful (Dawson \etal\ 2004), although the achieved depth
(\Heiiuv/Lya $<$ 13--20 \% at 2--3 $\sigma$ and $W($\Heiiuv$) <$
17--25 \AA) may not be sufficient.  The origin of these high $W(\lya)$
remains thus unclear.  

However, there is some doubt on the reality of the LALA high
equivalent widths measured from NB and broad-band imaging, or at least
on them being so numeours even at $z=4.5$. First of all the objects
with the highest $W(\lya)$ have very large uncertainties since the
continuum is faint or non-detected. Second, the determination of
$W(\lya)$ from a NB and a centered broad-band filter ($R$-band in the
case of Malhotra \& Rhoads 2002) may be quite uncertain, e.g.\ due to
unknowns in the continuum shape, the presence of a strong spectral
break within the broad-band filter etc.\ (see Hayes \& Oestlin 2006
for a quantification, and Shimasaku \etal\ 2006 ).  Furthermore other
groups have not found such high $W$ objects (Hu \etal\ 2004, Ajiki
\etal\ 2003) suggesting also that this may be related to insufficient
depth of the LALA photometry.

More recently larger samples of LAE were obtained, e.g.\ at $z=5.7$
(e.g.\ Shimasaku \etal\ 2006 has 28 spectroscopically confirmed
objects).  Although their {\em observed} restframe equivalent widths
$W^{\rm rest}_{\rm obs}(\lya)$ (median value and $W$ distribution) are
considerably lower than those of Malhotra \& Rhoads at $z=4.5$, and
only few objects (1--3 out of 34) show $W^{\rm rest}_{\rm obs}(\lya)
\ga 200$ \AA, it is possible that in several of these objects the
maximum \lya\ equivalent width of normal stellar populations is indeed
exceeded. This would clearly be the case if the IGM transmission at
this redshift is $T_\alpha \sim$ 0.3--0.5 (cf.\ Shimasaku \etal\ 2006,
Dijkstra \etal\ 2006c), which would imply that the true intrinsic
$W^{\rm rest} = 1/T_\alpha \times W^{\rm rest}_{\rm obs}$ is $\sim$ 2--3
times higher than the observed one. Shimasaku \etal\ estimate that
$\sim$ 30--40 \% of their LAE have $W^{\rm rest}(\lya) \ge$ 240 \AA\
and suggest that these may be young galaxies or again objects
with PopIII contribution. 
Dijkstra \& Wyithe (2007), based on \lya-LF and $W(\lya)$ modeling,
also argue for the presence of PopIII stars in this $z=5.7$ LAE sample.

Another interesting result is the increase of the fraction of large
$W(\lya)$ LBGs with redshift, e.g.\ from $\sim$ 2 \% of the objects
with $W^{\rm rest}(\lya) > 100$ \AA\ at $ \sim 3$ to $\sim$ 80 \%
at redshift 6, which is tentatively attributed  lower extinction,
younger ages or an IMF change (Shimasaku \etal\ 2006, Nagao \etal\ 2007).

Despite these uncertainties it is quite clear that several 
very strong LAE emitters are found and that these objects 
are probably the most promising candidates to detect direct {\em in situ}
signatures of PopIII at high redshift (see also Scannapieco \etal\ 2003).
Searches are therefore ongoing (e.g.\ Nagao \etal\ 2005) and the first
such discovery may be ``just around the corner'', or may need more
sensitive spectrographs and multi-object near-IR spectroscopy
(cf.\ Sect.\ \ref{s_future}). 

\subsubsection{Dust properties of high-$z$ LAE}
Although there are indications that LAE selected through their \lya\
emission are mostly young and relatively dust free objects 
(e.g.\ Shimasaku \etal\ 2006, Pirzkal \etal\ 2006, Gawiser \etal\ 2006),
% Overzier \etal\ 2006, 
%Finkelstein \etal\ 2006), 
it is of great interest to search for signatures of dust in
distant/primeval galaxies \footnote{Remember that e.g.\ sub-mm
  selected galaxies -- i.e.\ very dusty objects -- or at least a
  subsample of them show also \lya\ emission (Chapman \etal\
  2003).}.
Furthermore some models predict a fairly rapid production and the 
presence of significant amounts of dust at high-$z$ (Mao \etal\ 2006).
 LAE have the advantage of being at known redshift and of
indicating the presence of massive stars. SED fits of such objects
must therefore include populations of $< 10$ Myr age providing thus an
additional constraint on modeling.

\begin{figure}[tb]
%\centerline{\psfig{file=lai_etal2006_fig5,width=6cm}
%  \psfig{file=lai_etal2006_fig4.eps,width=6cm}}
\centerline{\psfig{file=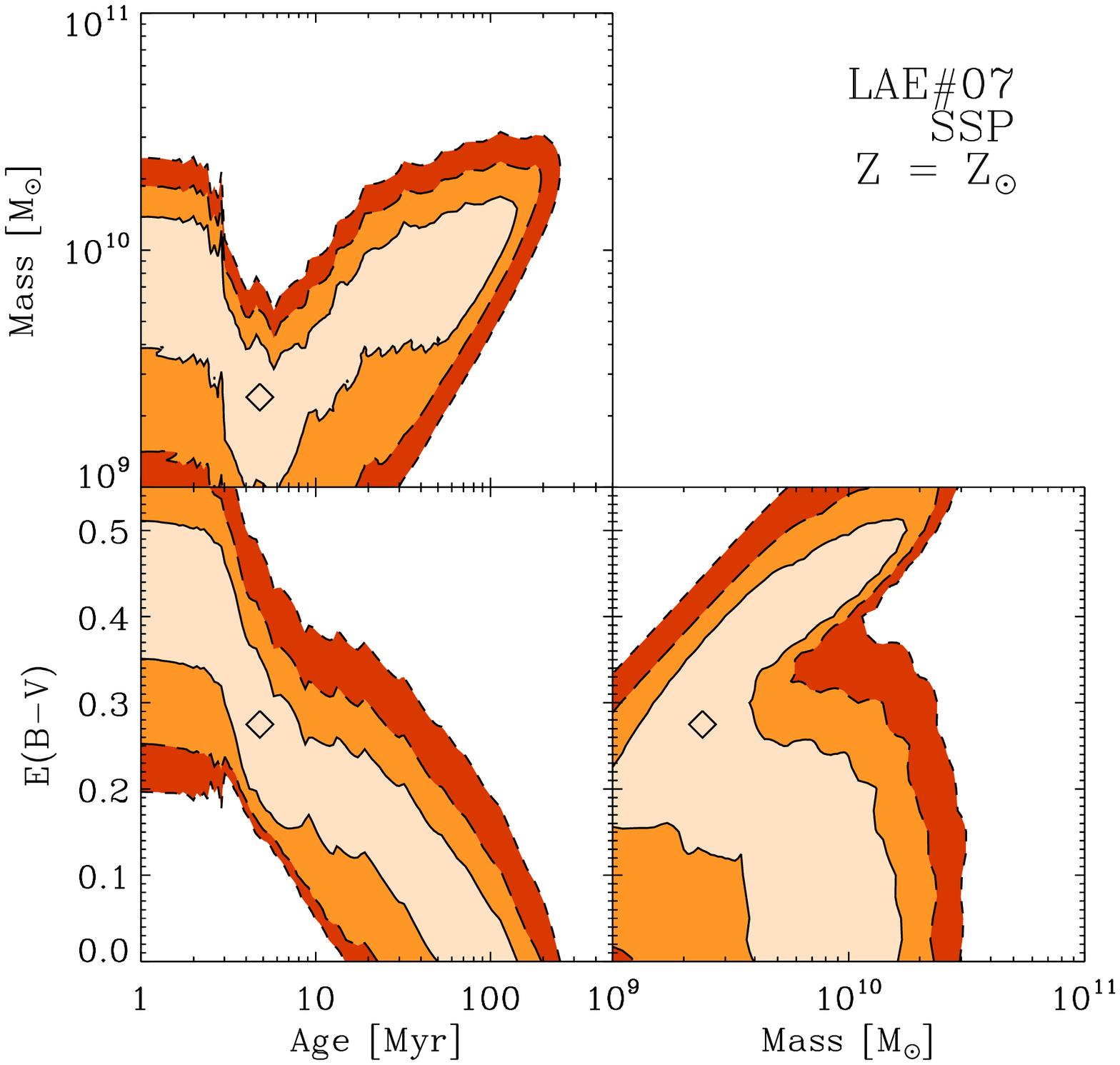,width=6cm}
  \psfig{file=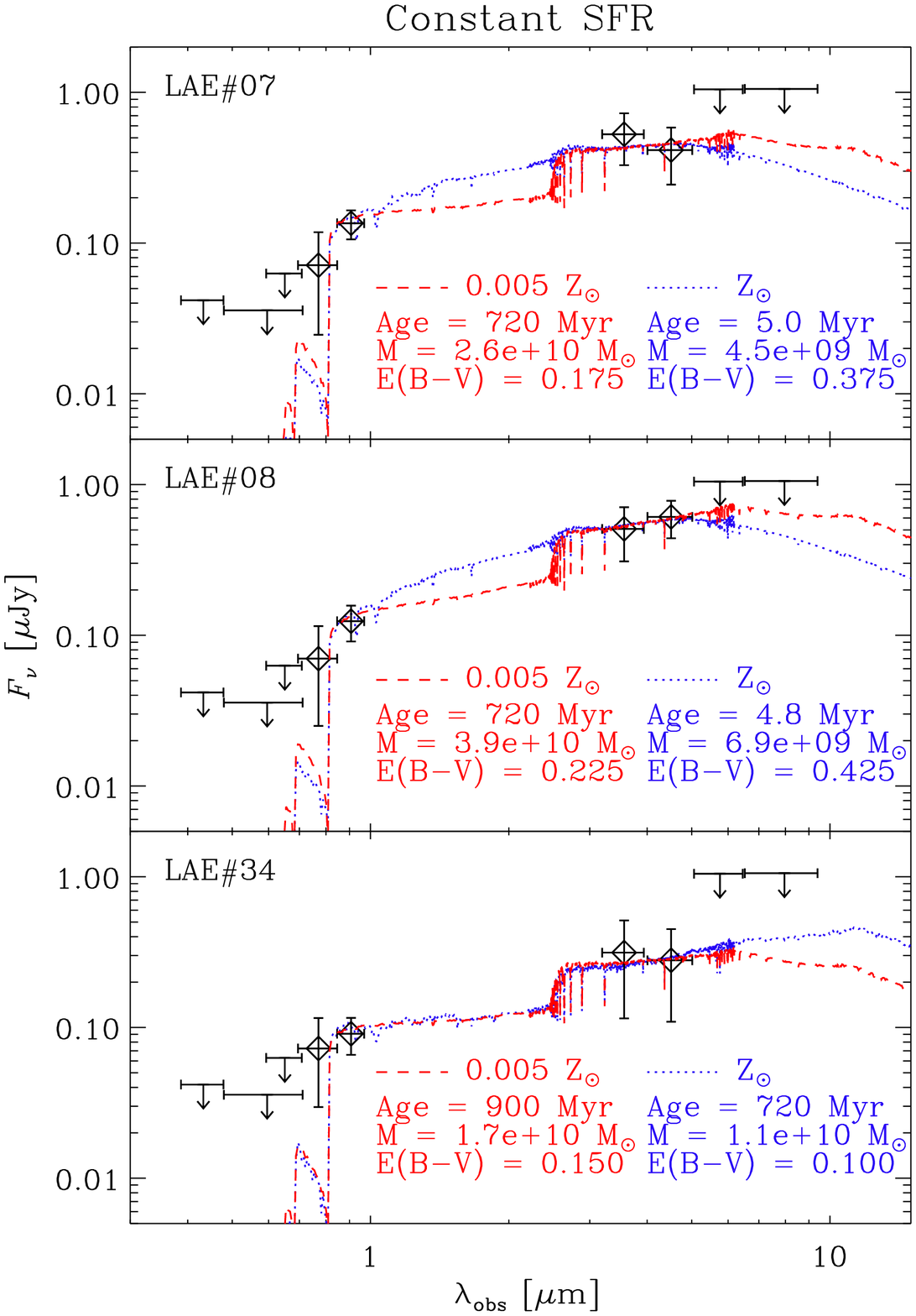,width=6cm}}
\caption{Illustrations of SED fits to $z=5.7$ LAE from Lai \etal\ (2007).
{\bf Left:} $\chi^2$ contour plots showing the best solution for one
object and degeneracies in the fit parameter. 
{\bf Right:} Comparison of best fit SEDs with constant SFR to observations
for 3 LAE.
These results show indications for the presence of dust in $z=5.7$ LAE.
See text for discussion.}
\label{fig_lai}
\end{figure}	

Recently the stellar populations of some high-$z$ LAEs have been analysed
with such objectives in mind.
For example the $z=6.56$ gravitationally lensed LAE discovered by Hu \etal\ (2002)
has recently been analysed by Schaerer \& Pell\'o (2005), who find that
a non-negligible extinction ($A_V \sim 1.$) may be necessary to 
reconcile the relatively red UV-restframe SED and the presence of \lya.
Later this interpretation was supported by Chary \etal\ (2005) including
also longer wavelength photometry obtained with Spitzer.
Three NB selected LAE at $z=5.7$ detected in the optical and with
Spitzer at 3.6 and 4.5 \micron, have recently been analysed by Lai
\etal\ (2007). Overall they find SED fits degenerate in age,
extinction, metallicity and SF history with stellar population ages up
to 700 Myr.  Most solutions require some dust extinction (see Fig.\
\ref{fig_lai}). If the need for \lya\ emission, i.e.\ for the presence
of young (massive) stars is taken into account, it seems that a
constant SFR scenario is likely together with an extinction of
$E_{B-V} \sim$ 0.1--0.2.

Although still uncertain, these four $z \sim$ 5.7--6.6 galaxies
provide currently to my knowledge the best indications for dust in
``normal'' galaxies around 1 Gyr after the Big Bang
\footnote{Dust emission has been observed in quasars out to
$z \sim 6$, as discussed briefly in Sect.\ \ref{s_dust}.}.
As already mentioned, these objects are probably not representative
of the typical high-$z$ LAE, but they may be of particular 
interest for direct searches of high-$z$ dust.
In any case, the first attempts undertaken so far to detect 
dust emission from $z \sim 6.5$ galaxies in the sub-mm (Webb \etal\ 2007,
Boone \etal\ 2007) have provided upper limits on their dust masses
of the order of $\sim (2-6) \times 10^8$ \msun.
Future observations with more sensitive instruments and targeting
gravitationally lensed objects should soon allow progress
in this field.

%%%%%%%%%%%%%%%%%%%%%%%%%%%%%%%%%%%%%%%%%%%%%%%%%%%%%%%%%%%%%%%%%%%%%%%%
%\section{Distant/primeval galaxies: Lyman-break galaxies}
\subsection{Lyman-break galaxies}
\label{s_lbg}

In general Lyman-break galaxies (LBGs) are better known than the
galaxies selected by \lya\ emission (LAE) discussed above.  There is a
vast literature on LBGs, summarised in an annual review paper in 2002
by Giavalisco (2002). However, progress being so fast in this area, frequent
``updates'' are necessary. In this last part I shall give an overview of the current
knowledge about LBGs at $z \ga$ 6, trying to present the main methods,
results, uncertainties and controversies, and finally to summarise the main 
open questions.
Are more general overview about galaxies across the Universe and out
to the highest redshift is given in the lectures of Ellis (2007).
Recent results from deep surveys including LBGs and LAE can be found
in the proceedings from ``At the Edge of the Universe'' (Afonso \etal\ 2007).
Giavalisco (this Winterschool) also covers in depth galaxy surveys.

The general principle of the LBG selection has already been mentioned above.
The number of galaxies identified so far is approximately:
4000 $z \sim 4$ galaxies (B-dropout), 1000 $z \sim 5$ galaxies (V-dropout),
and 500 $z \sim 6$ galaxies (i-dropout) according to the largest
dataset compiled by Bouwens and collaborators (cf.\ Bouwens \& Illingworth 2006).
The number of good candidates at $z \ga 7$ is still small (cf.\ below).

\subsubsection{i-dropout ($z \sim 6$) samples}
\label{s_z6}

Typically two different selections are applied to find $z \sim 6$
objects.  {\em 1)} a simple $(i-z)_{AB} >$ 1.3--1.5 criterium
establishing a spectral break plus optical non-detection, or {\ 2)}
$(i-z)_{AB} > 1.3$ plus a blue UV (restframe) slope to select actively
star forming galaxies at these redshift.
The main samples have been found thanks to deep HST imaging 
(e.g.\ in the Hubble Ultra-Deep Field and the GOODS survey),
and with SUBARU
(see Stanway \etal\ 2003, 2004, Bunker \etal\ 2004, Bouwens \etal\ 2003,
Yan \etal\ 2006)

In general all photometric selections must avoid possible
``contamination'' by other sources. For the $i$-dropouts possible contaminants
are: L or T-dwarfs, $z \sim$ 1--3 extremely red objects (ERO), or 
spurious detections in the $z$ band. 
Deep photometry in several bands (ideally as many as possible!) is required
to minimize the contamination.
The estimated contamination of $i$-drop samples constructed using
criterium 1) is somewhat controversial and could reach up to 25 \% in
GOODS data e.g., according to Bouwens \etal\ (2006) and Yan \etal\
(2006).
Follow-up spectroscopy has shown quite clearly that L-dwarfs contaminate 
the bright end of the $i$-dropout samples, whereas at fainter magnitudes
most objects appear to be truly at high-$z$ (Stanway \etal\ 2004,
Malhotra \etal\ 2005).

The luminosity function (LF) of $z \sim 6$ LBGs has been measured and
its redshift evolution studied by several groups.  Most groups
find an unchanged faint-end slope of $\alpha \sim -1.7$ from $z \sim$
3 to 6. Bouwens \etal\ (2006) find a turn-over at the bright end of
the LF, which they interpret as being due to hierarchical buildup of
galaxies.  However, the results on $M\star$ and $\alpha$ remain
controversial.  For example Sawicki \& Thompson (2006) find no change of
the bright end of the LF but an evolution of its faint end from $z
\sim 4$ to 2, while other groups (e.g.\ Bunker \etal\ 2004, Yoshida
\etal\ 2006, Shimasaku \etal\ 2006) find similar results as Bouwens
\etal.  The origin of these discrepancies remain to be clarified.

The luminosity density of LBGs and the corresponding star formation
rate density (SFRD) has been determined by many groups up to redshift
$\sim 6$.  Most of the time this is done by integration of the LF down to a
certain reference depth, e.g.\ $0.3 L_\star(z=3)$, and at high-$z$
generally no extinction corrections are applied. Towards high-$z$, the
SFRD is found to decrease somewhat from $z \sim 4$ to 6, whereas
beyond this the results are quite controversial as we will discuss
(see e.g.\ a recent update by Hopkins 2006).

The properties of individual galaxies will be discussed in Sect.\ \ref{s_indiv}.

\begin{figure}[tb]
%\centerline{\psfig{file=SFR_v9.eps,angle=270,width=10cm}}
\centerline{\psfig{file=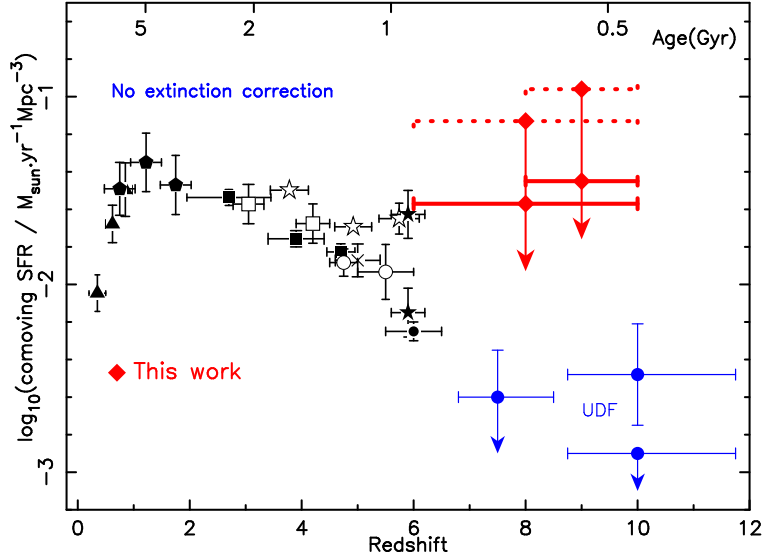,angle=270,width=10cm}}
\caption{
Evolution of the comoving 
  SFR density as a function of redshift including a compilation of
  results at $z \protect\la 6$, estimates from the lensing cluster
survey of Richard \etal\ (2006) 
 for the redshift ranges [$6-10$] and [$8-10$], and the
  values derived by Bouwens and collaborators from the Hubble
  Ultra-Deep Field (labeled ``UDF'').  Red solid lines: SFR density obtained
  from integrating the LF of the first category candidates of Richard \etal\ down to
  $L_{1500}=0.3\ L^{*}_{z=3}$; red dotted lines: same as red solid
  lines but including also second category candidates with a detection
  threshold of $<2.5 \sigma$ in $H$. From Richard \etal\ (2006).}
\label{fig_sfrd}
\end{figure}	

\subsubsection{Optical-dropout samples ($z \protect\ga 7$)}
\label{s_z7}
Going beyond redshift 7 requires the use of near-IR observations,
as the \lya-break of such objects moves out of the optical window.
Given the different characteristics of such detectors and imagers
(lower sensitivity and smaller field of view) progress has been
less rapid than for lower redshift observations.

In the NICMOS Ultra-Deep Field, Bouwens \etal\ (2004, 2006) have found
1--4 $z$-dropouts detected in $J$ and $K$, compatible with redshift 
$z \sim$ 7 starbursts.  
From this small number of objects and from the non-detection of $J$-dropouts
by Bouwens \etal\ (2005) they deduce a low SFR density between $z \sim$ 7 and 
10, corresponding corresponding to a significant decrease of the SFRD with
respect to lower redshift (see Fig.\ \ref{fig_sfrd}, symbols labeled ``UDF'').
The properties of these and other $z \ga 7$ galaxies will be discussed
below.

As an alternative to ``blank fields'' usually chosen for ``classical''
deep surveys, the use of gravitational lensing clusters -- i.e.\
galaxy clusters acting as strong gravitational lenses for background
sources -- has over the last decade or so proven very efficient in
finding distant galaxies (e.g.\ Hu \etal\ 2002, Kneib \etal\ 2004).
Using this method, and applying the Lyman-break technique plus a
selection for blue UV restframe spectra (i.e.\ starbursts), our group
has undertaken very deep near-IR imaging of several clusters to search
for $z \sim$ 6--10 galaxy candidates (see Schaerer \etal\ 2006 for an
overview).  13 candidates whose SED is compatible with that of star
forming galaxies at $z > 6$ have been found (see Richard \etal\ 2006
for detailed results). After taking into account the detailed lensing
geometry, sample incompleteness, correcting for false-positive
detections, and assuming a fixed slope taken from observations at $z
\sim 3$, their LF was computed.  Within the errors the resulting LF is
compatible with that of $z \sim 3$ Lyman break galaxies.  At low
luminosities it is also compatible with the LF derived by Bouwens
\etal\ (2006) for their sample of $z\sim 6$ candidates in the Hubble
Ultra Deep Field and related fields. However, the turnover observed by
these authors towards the bright end relative to the $z\sim 3$ LF is
not observed in the Richard \etal\ sample.
The UV SFR density at $z \sim$ 6--10 determined from this LF is shown
in Fig.\ \ref{fig_sfrd}.  These values indicate a similar SFR density
as between $z \sim$ 3 to 6, in contrast to the drop found from the
deep NICMOS fields (Bouwens \etal\ 2006)\footnote{The SFRD values of
  Bouwens have been revised upwards, reducing the differences with our
  study (see Hopkins 2007)}. The origin of these differences
concerning the LF and SFRD remain unclear until now.  In any case,
recent follow-up observations with HST and Spitzer undertaken to
better constrain the SEDs of the these candidates or to exclude some
of them as intermediate-$z$ contaminants, show that the bulk of our
candidates are compatible with being truly at high-$z$ (see Schaerer
\etal\ 2007a).

One of the main avenues to clarify these differences is by improving
the statistics, in particular by increasing the size (field of view)
of the surveys. Both surveys of more lensing clusters and wide blank
field near-IR surveys, such as UKIDSS are ongoing.
First $z \sim$ 7 candidates have recently been found by UKIDSS
(McLure 2007, private communication).

In this context it should also be remembered that not all optical
dropout galaxies are at high-$z$, as a simple ``dropout''
criterium only relies on a very red color between two adjacent filters.
As discussed for the $i$-dropouts above, extremely red objects
(such as ERO) at $z \sim$ 1--3 can be selected by such criteria.
See Dunlop \etal\ (2007) and Schaerer \etal\ (2007b) for such examples.
This warning is also of concern for searches for possible
massive (evolved) galaxies at high redshift as undertaken by
Mobasher \etal\ (2005) and McLure \etal\ (2006).

\begin{figure}[tb]
%\centerline{\psfig{file=plot_sed_7rev_all.eps,width=6cm}
%\psfig{file=labbe_etal2006_fig3.eps,width=6cm}}
\centerline{\psfig{file=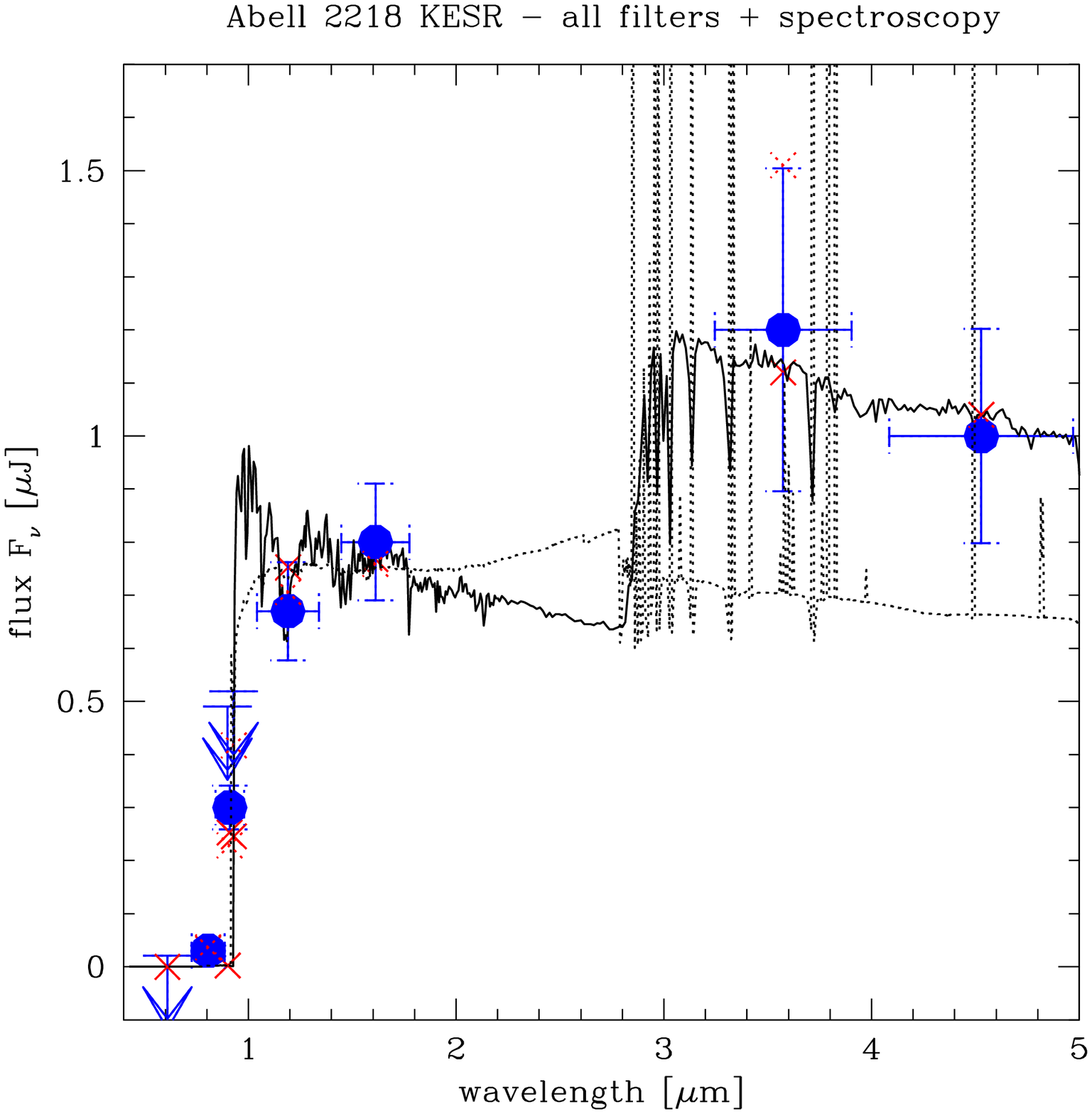,width=6cm}
\psfig{file=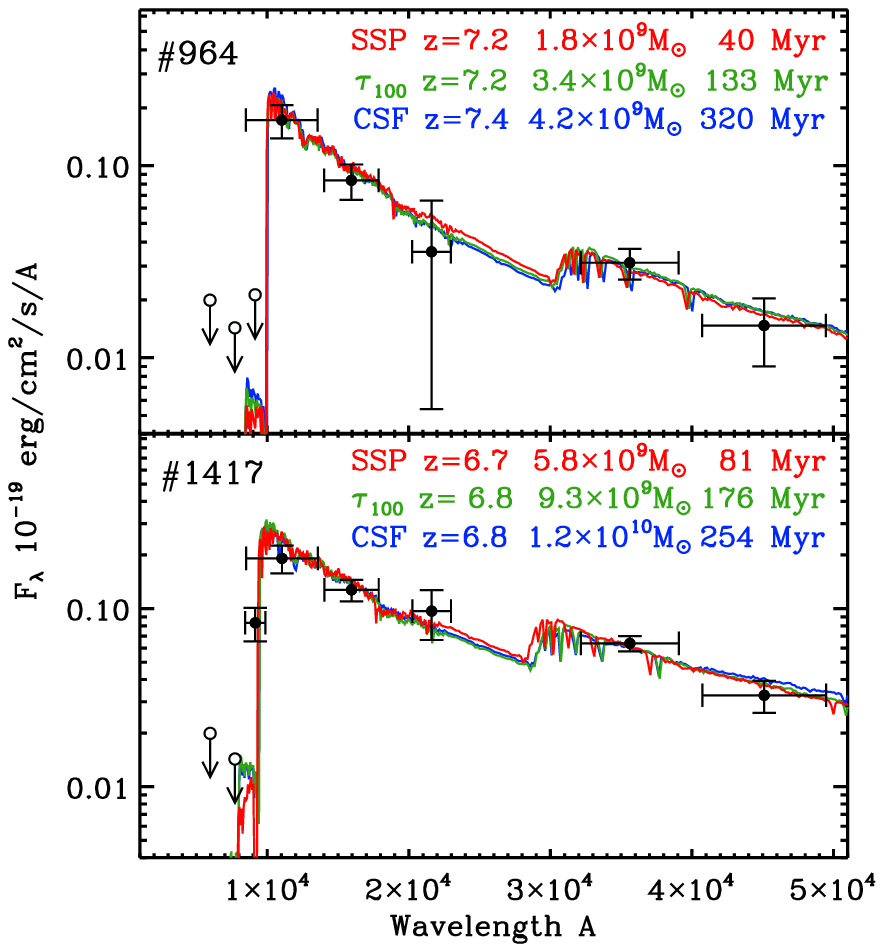,width=6cm}}
\caption{{\bf Left:} Observed SED of the $z \sim 7$ lensed galaxy from 
Egami \etal\ (2005) and model fits from Schaerer \& Pell\'o (2005) showing 
possible solutions with young ages ($\sim$ 15 Myr, solid line) or with 
a template of a metal-poor galaxy showing strong emission lines.
{\bf Right:} SEDs of two IRAC detected $z \sim 7$ galaxies from the Hubble
Ultra Deep field and best fits using different SF histories.
From Labb\'e \etal\ (2006). Note the different flux units ($F_\nu$ versus
$F_\lambda$) used in the two plots.}
\label{fig_seds}
\end{figure}	

\subsubsection{Properties of $z \protect\ga 6$ LBG galaxies}
\label{s_indiv}
Let us now review the main properties of individual $z \ga 6$ LBG,
i.e.\ continuum selected, galaxies and discuss implications thereof.
\lya\ emitters (LAE), such as the $z=6.56$ lensed galaxy found by
Hu \etal\ (2002), have already been discussed earlier (Sect.\ \ref{s_lae}).
Determinations of stellar populations (ages, SF history), extinction,
and related properties of such distant galaxies have really been possible
only recently with the advent of the Spitzer space telescope
providing sensitive enough imaging at 3.6 and 4.5 \micron.
These wavelengths, longward of the $K$-band and hence not
available for sensitive observations from the ground, correspond to the
restframe optical domain, which is crucial to constrain
properly stellar ages and stellar masses.
 
% % % % % % % A 2218
A triply lensed galaxy high-$z$ galaxy magnified by a factor $\sim 25$
by the cluster Abell 2218 has been found by Kneib \etal\ (2004).
Follow-up observations with Spitzer allowed to constrain its SED up to
4.5 \micron\ and show a significant Balmer break (Egami \etal\ 2005,
see Fig.\ \ref{fig_seds}).  Their analysis suggests that this $z \sim
7$ galaxy is in the poststarburst stage with an age of at least $\sim
50$ Myr, possibly a few hundred million years. If true this would
indicate that a mature stellar population is already in place at such
a high redshift.  However, the apparent 4000-\AA\ break can also be
reproduced equally well with a template of a young ($\sim 3-5$ Myr)
burst, where strong rest-frame optical emission lines enhance the 3.6-
and 4.5\micron\ fluxes (Schaerer \& Pell\'o 2005, and Fig.\
\ref{fig_seds}).  The stellar mass is an order of magnitude smaller
($\sim 10^9$ \msun) smaller than that of typical LBG, the extinction
low, and its SFR $\sim$ 1 \msunyr.
% % % % %

Two to four of the four $z \sim 7$ candidates of Bouwens \etal\ (2004)
discussed above
have been detected in the very deep 23.3h exposures taken with Spitzer
at 3.6 and 4.5 \micron\ by Labb\'e \etal\ (2006). Their SED analysis
indicates photometric redshifts in the range 6.7-7.4, stellar masses
$(1-10)\times 10^9$ \msun, stellar ages of 50--200 Myr, and star
formation rates up to $\sim 25$ \msunyr, and low reddening $A_V<0.4$.

\begin{figure}[tb]
%\centerline{%\psfig{file=plot_sed_7rev_all.eps,width=6cm}
%\psfig{file=eylesl_fig19.ps,width=10cm}}
\centerline{\psfig{file=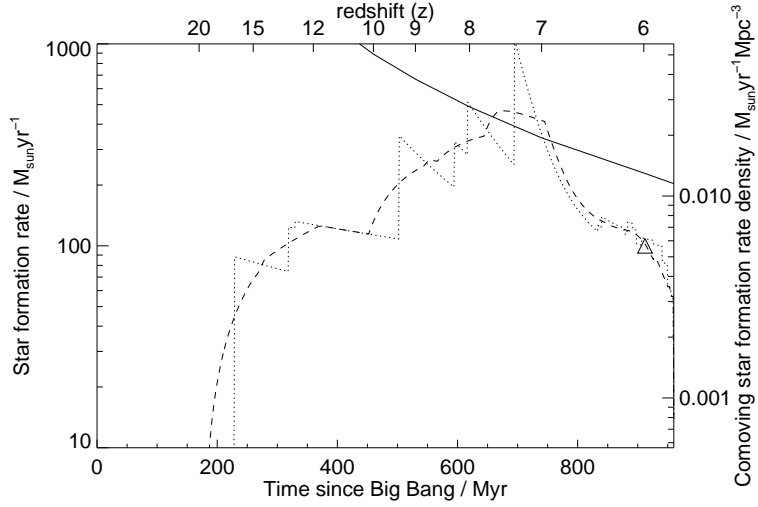,width=10cm}}
\caption{History of the star formation rate density determined by inversion
from the observed $i$-dropout galaxies analysed by Eyles \etal\ (2007).
The dotted curve is the sum of the past
star formation rates for our $i'$-drop sample (left axis, with the corresponding
star formation rate density shown on the right axis, corrected for incompleteness
including a factor of 3.2 for galaxies below the flux threshold.
The dashed curve is this star formation history smoothed on a timescale of 100\,Myr.
The triangle is the estimate of the unobscured (rest-frame UV) star formation rate density
at $z\approx 6$ from $i'$-drops in the HUDF from Bunker et al.\ (2004).
The solid curve shows the condition for reionisation from star formation,
as a function of time (bottom axis) and redshift (top axis),
assuming an escape fraction of unity for the Lyman continuum photons. 
From Eyles \etal\ (2007).}
\label{fig_eyles}
\end{figure}

Evidence for mature stellar populations at $z \sim 6$ has also been
found by Eyles \etal\ (2005, 2007). By ``mature'' or ``old'' we mean
here populations with ages corresponding to a significant fraction of
the Hubble time, which is just $\sim$ 1 Gyr at this redshift.
Combining HST and Spitzer data from the GOODS survey they find that 40
\% of 16 objects with clean photometry have evidence for substantial
Balmer/4000-Å spectral breaks. For these objects, they find ages of
$\sim$ 200--700 Myr, implying formation redshifts of $7 \le z_f \le
18$, and large stellar masses in the range $\sim (1-3) \times ×
10^{10}$ \msun.  Inverting the SF histories of these objects they
suggest that the past global star formation rate may have been much
higher than that observed at the $z \sim 6$ epoch, as shown in Fig.\ 
\ref{fig_eyles}. This could support
the finding of a relatively high SFR density at $z \ga 7$, such as
found by Richard \etal\ (2006).

In short, although the samples of $z > 6$ Lyman break galaxies for
which detailed information is available are still very small, several
interesting results concerning their properties emerge already: mature
stellar populations in possibly many galaxies indicating a high formation
redshift, stellar masses of the order of $10^9$ to $10^{10}$ \msun,
and generally low extinction.
However a fraction of these galaxies 
%-- the ones which remain undetec
appears also to be young and less massive (cf.\ Eyles \etal\ 2007)
forming a different ``group''.
Similar properties and similar two groups are also found among the
high-$z$ LAE (cf.\ Schaerer \& Pell\'o 2005, Lai \etal\ 2007, and
Pirzkal \etal\ 2006) already discussed above.  Whether such separate
``groups'' really exist and if so why, or if there is a continuity of
properties remains to be seen.

In a recent analysis Verma \etal\ (2007) find that
$\sim$ 70 \% of $z \sim 5$ LBGs have typical ages of $\la$ 100 Myr,
and stellar masses of $\sim 10^9$ \msun, which are younger and less
massive than typical LBGs at $z \sim 3$.
They also find indications for a relatively low extinction, lower than
at $z \sim 3$.
The trend of a decreasing extinction in LBGs with increasing redshift
has been found by many studies, and is in agreement with the results
discussed above for $z \sim 6$ and higher.
However, the differences in age and mass e.g.\ compared with the 
objects of Eyles \etal\ (2007) may be surprising, especially
given the short time ($\sim$ 200 Myr) between redshift 5 and 6.
Several factors, such as selection effects, the representativity
of the small $z \sim 6$ samples studied in detail, etc.\
may contribute to such differences. 
Reaching a more complete and coherent understanding of the different
primeval galaxy types, of their evolution, and their relation with 
galaxies at lower redshift will need more time and further observations.

%\subsubsection{Connections with LAE}
%\label{s_lbglae}
%Mao, Lapi \etal\ 
%Pentericci \etal\ 2007 (astro-ph/0703013)
%
%Shimasaku \etal\ (2006) 

%%%%%%%%%%%%%%%%%%%%%%%%%%%%%%%%%%%%%%%%%%%%%%%%%%%%%%%%%%%%%%%%%%%%%%%%
\subsection{What next?}
\label{s_future}
%\subsubsection{Open questions}

As shown in these lectures it has been possible during the last decade 
to push the observational limits out to very high redshift and to 
identify and to study the first samples of galaxies observed barely 
$\sim$ 1 Gyr after the Big Bang.
The current limit is approximately at $z \sim$ 7--10, where just few
galaxies (or galaxy candidates) have been detected, and where spectroscopic
confirmation remains extremely challenging.

Thanks to very deep imaging in the near-IR domain it is possible to
estimate or constrain the stellar populations (age, SF history, mass, etc.)
and dust properties (extinction) of such ``primeval'' galaxies,
providing us with a first glimpse on galaxies in the early universe.
Despite this great progress and these exciting results, the global 
observational picture on primeval galaxies, their formation and evolution,
remains to be drawn. Many important questions remain or, better said,
start to be posed now, and can now or in the near future be addressed 
not only by theory and modeling but also observationally!

We have already seen some of the emerging questions in several parts 
of these lecture. Others, sometimes more general ones, have not been
addressed.  Among the important questions concerning primeval galaxies 
we can list:
\begin{itemize}
\item How do different high-$z$ populations such as LAE and LBG fit together? Are
there other currently unknown populations? What are the evolutionary links between
these populations and galaxies at lower redshift?
\item What is the metallicity of the high-$z$ galaxies? Where is Population III? 

\item What is the star formation history of the universe during the first Gyr
after Big Bang?
\item Are there dusty galaxies at $z \ga 6$? How, where, when, and how much dust is produced
at high redshift?
\item Which are the sources of reionisation?  And, are these currently detectable
galaxies or very faint low mass objects? What is the history of cosmic reionisation?
\end{itemize}

We, and especially you the young students, are fortunate to live in a period
where theory, computing power, and many future observational facilities
are rapidly growing, enabling astronomers to peer even deeper into the universe.
It is probably a fair guess to say that within the next 10-20 years we
should have observed the truly first galaxies forming in the universe,
found Population III, etc.\ 
We will thus have reached the limits of the map in this exploration
of the universe. However, a lot of challenging and interesting work
will remain to reach a global and detailed understanding of the formation
and evolution of stars and galaxies!

%%%%%%%%%%%%%%%%%%%%%%%%%%%%%%%%%%%%%%%%%%%%%%%%%%%%%%%%%%%%%%%%%%%%%%%%
%\noindent
%{\bf Acknowledgements}
\subsection*{Acknowledgements}

I thank Jordi Cepa for the invitation to lecture on this topic
and for his patience with the manuscript. I would also like to thank
him and the IAC team for the organisation of this excellent and
enjoyable winterschool.
Over the last years I've appreciated many interesting and stimulating discussions 
with my collaborators and other colleagues. Among them I'd like to thank
in particular
Roser Pell\'o, Johan Richard, Jean-Fran\c{c}ois Le Borgne, Jean-Paul Kneib,
Angela Hempel, and Eiichi Egami representing the ``high''-$z$ universe,
Daniel Kunth, Anne Verhamme, Hakim Atek, Matthew Hayes, and Miguel Mas-Hesse
for the nearby universe, as well as
Andrea Ferrara, Grazyna Stasi\'nska, and David Valls-Gabaud.
Both the list of people to thank and the literature list is quite incomplete
though. Apologies.

%%%%%%%%%%%%%%%%%%%%%%%%%%%%%%%%%%%%%%%%%%%%%%%%%%%%%%%%%%%%%%%%%%%%%%%%
%\input{qq}
%\end{document}

%%%%%%%%%%%%%%%%%%%%%%%%%%%%%%%%%%%%%%%%%%%%%%%%%%%%%%%%%%%%%%%%%%%%%%%%%

%%%%%%%%%%%%%%%%%%%%%%%%%%%%%%%%%%%%%%%%%%%%%%%%%%%%%%%%%%%%%%%%%%%%%%%%
\end{document}